\documentclass[a4paper,fleqn]{cas-sc}

\usepackage[numbers]{natbib}

\PassOptionsToPackage{table}{xcolor}
\usepackage{framed}
\definecolor{shadecolor}{gray}{0.95}

\usepackage{soul}

\usepackage{xcolor}

\usepackage{tabularx}
\usepackage{colortbl}
\usepackage{tcolorbox}

\usepackage{hyperref}

\usepackage{arydshln}

\def\tsc#1{\csdef{#1}{\textsc{\lowercase{#1}}\xspace}}
\tsc{WGM}
\tsc{QE}

\begin{document}
\let\WriteBookmarks\relax
\def\floatpagepagefraction{1}
\def\textpagefraction{.001}

\shorttitle{Measuring Code Efficiency Optimization Capabilities with ACEOB}

\shortauthors{Yue Pan,Xiuting Shao, Chen Lyu}

\title [mode = title]{Measuring Code Efficiency Optimization Capabilities with ACEOB}



%


\author[SDNU1]{Yue Pan}\ead{1512132224@qq.com}


\author[SDNU1]{Xiuting Shao} \ead{shaoxiuting@126.com}

\author[SDNU1]{Chen Lyu\corref{cor1}} \ead{lvchen@sdnu.edu.cn}
\cortext[cor1]{Corresponding author}

\affiliation[SDNU1]{organization={School of Information Science and Engineering, Shandong Normal University},
            city={Jinan},
            postcode={250301}, 
            state={Shandong},
            country={China}}


















\begin{abstract}
As Moore's Law gains diminish, software performance and efficiency become increasingly vital. Optimizing code efficiency is challenging, even for professional programmers. However, related research remains relatively scarce, and rigorously assessing models' abilities to optimize code efficiency is fraught with difficulties. In response to this challenge, we first conduct an in-depth analysis of “code patterns” in the model training dataset, meticulously exploring human-written code. Secondly, we define a task for optimizing code efficiency and introduce the \textbf{A}utomatic \textbf{C}ode \textbf{E}fficiency \textbf{O}ptimization \textbf{B}enchmark (ACEOB), which consists of 95,359 pairs of efficient-inefficient code aimed at assessing code efficiency optimization capabilities. To our knowledge, ACEOB is the first dataset specifically targeting Python code efficiency optimization. To evaluate models' ability in optimizing code efficiency, we propose two new metrics: the \textbf{I}somorphic \textbf{O}ptimal \textbf{C}omparison \textbf{C}ode\textbf{B}LEU (IOCCB) metric and the \textbf{N}ormalized \textbf{P}erformance \textbf{I}ndex (NPI) metric, to assess the efficiency of model-generated code. We also evaluate several advanced code models, such as PolyCoder and CodeT5, after fine-tuning them on ACEOB and demonstrate that the efficiency of each model improves after introducing the NPI filter. However, it was observed that even ChatGPT does not perform optimally in code efficiency optimization tasks.
\end{abstract}


\maketitle



\section{Introduction}\label{1}

\subsection{Challenges and Current State of Code Efficiency}\label{1.1}

For novice programmers lacking deep knowledge of algorithmic efficiency, crafting code that is both \textbf{accurate} and \textbf{efficient} has been a significant challenge in the fields of Software Engineering (SE), Programming Languages (PL), and Artificial Intelligence (AI). Achieving this goal holds profound implications for reducing computational resource consumption. Imagine a scenario: a software company, particularly a gaming company, wishes to lower energy costs and enhance operational efficiency. By writing the correct code, the machine can automatically optimize the company's code, swiftly providing a  functionally identical yet more efficient version. This process eliminates the need for costly outsourcing to optimize application efficiency and can be achieved through interaction with machines [\citealp{zan2022neural}].

In the realm of software development, the operational efficiency of code is  \textbf{paramount}. Currently, the software industry's reliance on \textbf{Moore's Law} for automatic performance enhancement is slowing. Moreover, Sam Altman's proposition that “intelligence in the universe doubles every 18 months” implies an exponential increase in intelligent demand and, consequently, computing power. Therefore, with the deceleration of automatic performance enhancement and the doubling of demand for computing power,  reducing computational costs and carbon footprint has become an \textbf{urgent} task, making  writing high-performance code a pressing endeavor [\mbox{\citealp{chen2022learning}}].

Traditionally, improving code efficiency has relied on  \textbf{optimization compilers}, which repeatedly apply optimization techniques in the intermediate steps of generating machine code. For instance, Cummins et al. [\mbox{\citealp{cummins2021programl}}] proposed a portable, language-agnostic program graph representation originating from compiler intermediate representations (IRs). However, reliance on optimization compilers typically means no alteration to the source code, overlooking the possibility of directly optimizing the source code to preserve its semantics.

When optimizing source code, a key issue is how to achieve the desired functionality through enhanced code efficiency. Notably, achieving the same functionality can be accomplished through  \textbf{various inefficient and efficient implementations}. For example, loops can be written using recursion, for loops, while loops, etc., and the entire program's algorithmic structure may employ different methods. The execution time of these different solutions can vary significantly, usually due to a variety of factors such as algorithmic complexity, choice of data structures, and library usage [\citealp{chen2022learning}]. Therefore, the task of optimizing code efficiency itself is highly diverse and complex. To advance optimization techniques, we need to systematically classify and analyze optimization methods. It is for this reason we propose the Initiative for Code Efficiency Optimization (\textbf{I}nefficient \textbf{C}ode \textbf{to} \textbf{E}fficient \textbf{C}ode, abbreviated as \textbf{IC2EC}) task definition (see Section \ref{4}).

Optimizing the runtime efficiency of source code is a  challenging task, even for experienced programmers. This is because it requires an in-depth understanding of algorithmic complexity, coding patterns, and the complexity of interactions with hardware. While compilers can automatically optimize low-level performance, identifying higher-level optimizations remains a challenge, such as finding more effective solutions to the same problem. To date, these types of optimizations still largely depend on  \textbf{manual identification}.

In the field of code efficiency optimization research, the assessment metrics for evaluating the capability of \textbf{L}arge-scale \textbf{L}anguage \textbf{M}odel\textbf{s} (LLMs) in executing the Initiative for IC2EC task are still insufficient. Time complexity, as a universally adopted metric, is used to measure the efficiency of a given algorithm or code implementation. However, determining the worst-case time complexity of algorithms is often intractable [\citealp{turing1938computable}]. Thus, when employing static code analysis methods, including cyclomatic complexity [\citealp{mccabe1976complexity}], coupling, and Master Theorem analysis [\citealp{bentley1980general}], we must acknowledge these theoretical limitations. Meanwhile, many researchers have turned to dynamic code analysis to assess code complexity. This method is based on actual code execution and evaluates code execution time and space occupancy through a variety of test cases [\citealp{burnim2009wise, hutter2014algorithm,  nogueira2012predicting}]. Although dynamic analysis can effectively detect code errors, it relies on generating appropriate test cases and actual code execution, hence it has inherent limitations.

\textbf{Execution time} and \textbf{acceleration}, as traditional metrics, are widely applied under ideal circumstances—where all generated code is functionally correct and performance can be directly measured through execution. However, current code generation models are far from this ideal, resulting in assessments based on direct code execution that cannot accurately distinguish between a model's code \textbf{generation capacity} and \textbf{efficiency optimization ability}. For example, Model A might generate functionally correct code with 80\% accuracy, of which 20\% is highly efficient; whereas Model B might have only 20\% accuracy for generating functionally correct code but has an 80\% chance of generating more efficient code. Solely relying on execution time for assessment might lead to the misconception that both models have the same efficiency optimization capability, overlooking Model B's potential advantage in selecting efficient algorithms. This issue underscores that existing evaluation metrics cannot independently assess a model's ability to generate efficient code, but can only provide a comprehensive assessment of its ability to generate both functionally correct and efficient code.

Furthermore, in the domain of code efficiency optimization, researchers' exploration into code cost models remains relatively limited. For instance, Adams et al. [\citealp{adams2019learning}] utilized small neural networks to predict the execution time of Halide programs. Baghdadi et al. [\citealp{baghdadi2021deep}] proposed a cost model for predicting execution time for Tiramisu programs. However, these studies mainly focus on the areas of algorithmic base-level acceleration and compiler optimization, and currently, there is no research on cost models for high-level programming languages such as Python.

\subsection{The Need for Code Efficiency Optimization}\label{1.2}

As LLMs continue to advance, the concept of “\textbf{AI-assisted programming}” is becoming increasingly widespread. For instance, GitHub Copilot, an AI coding assistant based on CodeX [\citealp{chen2021evaluating}], aids developers by autogenerating code. AlphaCode [\citealp{li2022competition}], on the other hand, is focused on solving competitive programming challenges and has demonstrated performance comparable to human programmers. With the emergence of ChatGPT, providing programming assistance, error troubleshooting, and code optimization suggestions, AI-assisted programming tools have reached new heights.

However, current LLMs often generate code that is \textbf{less efficient}. For example, a significant amount of inefficient code is found in the outputs of CodeX, especially when tackling more complex problems. Similarly, AlphaCode faces the issue of “slow positive,” where, despite ensuring functional correctness, the generated code is inefficient and unable to meet time constraints. Lertbanjongam et al. [\citealp{lertbanjongngam2022empirical}] have analyzed the code generated by the AlphaCode system from an execution efficiency perspective, revealing that it performs worse in terms of execution time and memory usage compared to human-written code. The “slow positive” problem is particularly pronounced when dealing with complex issues, due to the overuse of nested loops and unnecessary variable declarations.

Notably, research on achieving a true “\textbf{programming parsing AI}” remains in its early stages. Ideally, if AI could parse and write efficient algorithmic programs like a human programmer, rather than merely memorizing “\textbf{code patterns}” from training datasets, it would be a significant breakthrough. Therefore, to delve deeper into the nature of AI learning, we need thorough statistical and research analysis of the “code patterns” seen by AI (see Section \ref{3}).

In this paper, we investigate the reasons why current code generation models fail to produce efficient code, summarized into the following three main aspects:
\begin{itemize}
\item \textbf{Programming Habit Influence:} We conducted statistics on approximately nine hundred thousand problem codes on the \mbox{\href{https://codeforces.com}{Codeforces}} website [\mbox{\citealp{codeforces}}] (see Section \mbox{\ref{3}}) and found that human programmers on the site tend to write inefficient code, using inefficient algorithms to solve problems. This tendency continues in pretrained models trained on human-written codes. As the difficulty of problems increases, the proportion of inefficient codes correspondingly rises, reflecting a prevalent efficiency issue in the “code patterns” written by humans.
\item \textbf{Lack of Efficiency-Related Datasets:} Existing code language models rarely have the opportunity to learn from comparisons between efficient and inefficient codes, lacking specific pretraining on code efficiency. Therefore, they cannot effectively distinguish the efficiency of different algorithmic solutions. The “slow positive” phenomenon observed in experiments with CodeX and AlphaCode indicates that the absence of relevant datasets prevents LLMs from effectively learning the intrinsic connection between efficient and inefficient codes.
\item \textbf{Lack of Effective Evaluation Metrics and Cost Models:} The current code evaluation metrics mainly focus on the functionality of the code, while relatively less attention is paid to code efficiency. This tendency leads to LLMs often neglecting efficiency factors in the code generation process, resulting in inefficient code that does not receive corresponding negative evaluation. In terms of assessing code efficiency, traditional metrics such as execution time and acceleration are usually based on the compilation and execution of the code. This approach assesses the functionality and efficiency of the code as an indivisible whole. However, given that the code generated by current LLMs often falls short in functionality, this fact further diminishes the emphasis on the efficiency of the generated code. To address this issue, we propose a new evaluation method that can assess code efficiency independently without the need for compilation, thus more comprehensively considering the overall performance of the code.
\end{itemize}

\subsection{Towards AI-Driven Code Efficiency Optimization: Proposals and Contributions}\label{1.3}

In this paper, we have chosen \textbf{programming competition websites} as the data source for the Initiative for IC2EC task to meet specific requirements for the needed dataset. Solution codes in programming competitions, while functionally identical, exhibit significant efficiency disparities (reflected in execution time) due to the use of different algorithms, making these datasets an ideal venue for assessing the capability of the IC2EC task.

We defined the IC2EC task (see Section \mbox{\ref{4}}) and introduced the \textbf{A}utomatic \textbf{C}ode \textbf{E}fficiency \textbf{O}ptimization \textbf{B}enchmark (\textbf{ACEOB}) benchmark dataset (see Section \mbox{\ref{5}}). With the rapid advancement of LLMs and their potential application in code efficiency optimization, ACEOB aims to comprehensively assess LLMs' performance on the IC2EC task. Additionally, we have released the \textbf{A}utomatic \textbf{C}ode \textbf{E}fficiency \textbf{O}ptimization \textbf{B}enchmark \textbf{Ori}ginal (\textbf{ACEOB-Ori}) and \textbf{A}utomatic \textbf{C}ode \textbf{E}fficiency \textbf{O}ptimization \textbf{B}enchmark\textbf{-}\textbf{N}ormalized \textbf{P}erformance \textbf{I}ndex (\textbf{ACEOB-NPI}) datasets for providing a reference of original data and training cost models, respectively.

The ACEOB benchmark offers a comprehensive and accurate method to measure the IC2EC task. It contains 95,359 pairs of \textbf{efficient-inefficient code pairs}, each pair intended to guide the model to learn specific efficiency optimization strategies. This benchmark covers 19 problem difficulties and 36 algorithm tags. If a model performs well on ACEOB, it indicates that it has mastered various algorithms and code optimization techniques and can flexibly use data structures and programming techniques.

Given that the metrics for assessing LLMs' performance on the IC2EC task are still insufficient, we introduced the \mbox{\textbf{I}somorphic \textbf{O}ptimal \textbf{C}omparison \textbf{C}ode\textbf{B}LEU (\textbf{IOCCB}) metric (see Section \ref{7.3})} and the \textbf{N}ormalized \textbf{P}erformance \textbf{I}ndex (\textbf{NPI}) metric (see Section \mbox{\ref{7.4}).} IOCCB is a variant of the CodeBLEU metric, specifically designed for the IC2EC task. Each item in the ACEOB dataset also includes alternate efficient codes, i.e., efficient codes with different algorithmic solutions, providing important references for IOCCB. Therefore, as long as the model generates sufficiently efficient code, it can score high even if it deviates from the true code. The NPI score is based on the calculation of the maximum/minimum execution time of codes that achieve the same functionality. These two metrics together provide a comprehensive quality assessment standard for the IC2EC task.

To address the inadequacy of Python code cost models, we developed two cost models: predicting Python code execution time and predicting Python code NPI scores. We have also made public the ACEOB-NPI dataset used for training these models (see Section \ref{6.5}).

We also developed a dedicated system for the IC2EC task and tested the performance of various mainstream code language models on the ACEOB benchmark after fine-tuning. Especially after introducing our \textbf{NPI filter}, the IC2EC task capability significantly improved, particularly when using sampling methods. In addition, we evaluated ChatGPT's [\citealp{chatgpt}] performance on the IC2EC task. Although it shows excellence in analyzing code complexity, our statistical analysis indicates that ChatGPT still faces challenges in generating efficient code.

\textbf{Contributions.} Our contribution falls into the following aspects:
\begin{itemize}
\item \textbf{Dataset.} We developed the ACEOB benchmark dataset, currently the first targeted at competition-level Python code efficiency optimization. We have also made public the ACEOB-Ori and ACEOB-NPI datasets.
\item \textbf{Motivation.} We analyzed the “code patterns” AI learns and the reasons behind generating inefficient code.
\item \textbf{Metrics \& Cost Models.} We introduced the IOCCB (see Section \ref{7.3}) and NPI (see Section \ref{7.4}) evaluation metrics and developed cost models for predicting Python code execution time and Python code NPI scores.
\item \textbf{Experimentation.} We tested the performance of mainstream code models on the IC2EC task and demonstrated the significant effect of the NPI filter in enhancing model performance.
\end{itemize}

Through the application of the ACEOB benchmark, we found that despite the continuous improvement in code generation performance, modern language models still face significant challenges in code efficiency optimization. With the rapid development of AI technology, the societal importance of code efficiency will continue to rise. Therefore, our proposed new benchmark provides an important reference standard for assessing LLMs' capabilities in the code efficiency optimization domain. This research is a critical step toward the automatic generation of high-performance code, fundamentally enhancing the performance of current developer tools and potentially reducing the carbon footprint of computational resources.

\section{Related Work}\label{2}

\subsection{Program Synthesis}\label{2.1}

Program synthesis focuses on \textbf{automatically generating programs} that meet given specifications. Task specifications may appear as \textbf{N}atural \textbf{L}anguage (\textbf{NL}) descriptions, a set of I/O examples, or constraints. Gulwani et al. [\citealp{gulwani2017program}] presented an excellent report on the adaptability of program synthesis methods. Deductive program synthesis defines a search problem using formal logic specifications and generates programs that meet these specifications, requiring complex optimization techniques [\citealp{alur2017sygus}]. A wide range of I/O examples  provide a method to generate programs that satisfy specific I/O examples [\citealp{cai2017making}], but this may not always produce the correct program. Using NL to describe program behavior is an alternative to formal or inductive specifications.

In recent years, advances in deep learning have spurred attempts to use it in the direction of program synthesis tools. For example, Yin and Neubig [\citealp{yin2017syntactic}] connected text and code to the \textbf{A}bstract \textbf{S}yntax \textbf{T}ree (\textbf{AST}) through the attention mechanism. Guo et al. [\citealp{guo2021learning}] focused on generating program sketches, and Devlin et al.'s [\citealp{devlin2017robustfill}] work on directly generating target programs. Moreover, there are many using deep learning methods [\citealp{balog2016deepcoder, bunel2018leveraging, kalyan2018neural, devlin2017neural, lee2018accelerating, nye2019learning, odena2020learning, pan2023measuring, parisotto2016neuro}], including neural [\citealp{zhao2018neural}] and non-neural [\citealp{meng2011systematic}] network models for editing code, but they are used for different applications, such as fixing errors [\citealp{yasunaga2020graph, chen2021plur}]. These studies have advanced the application of deep learning in the field of program synthesis, but they typically focus on writing functional code without considering the efficiency of the code.

One research direction for analyzing program efficiency is \textbf{analyzing time complexity}. McCabe et al. [\citealp{mccabe1976complexity}] introduced cyclomatic complexity, a metric for analyzing program complexity. Bentley et al. [\citealp{bentley1980general}] proposed divide-and-conquer algorithms, which describe the time complexity of programs as recursive relations, while Sikka et al. [\citealp{sikka2020learning}] proposed a learning-based code complexity prediction method. Narayanan et al. [\citealp{ narayanan2017graph2vec}] applied the neural graph embedding framework to the program's AST. Recently, Prenner and Robbes [\citealp{prenner2021making}] proposed using pre-trained code language models (such as CodeBERT [\citealp{feng2020codebert}]) for code complexity prediction, while Jeon et al.'s [\citealp{jeondeep}] hierarchical architecture and pre-training objectives further improved the accuracy of complexity prediction. Our research, distinct from these works, focuses on optimizing code efficiency.

\subsection{Automatic Code Optimization in Compilers}\label{2.2}

\textbf{Automatic code optimization} in compilers is one of the oldest problems in the field of compilers, aiming to develop methods for compilers that can automatically optimize code. Ashouri et al. [\citealp{ashouri2018survey}] reviewed the application progress of machine learning in the field of compiler optimization. In terms of frameworks, Cummins et al. introduced CompilerGym [\citealp{cummins2022compilergym}], an environment that conceives compiler optimization problems as easy-to-use. Ansel et al. [\citealp{ansel2014opentuner}] proposed OpenTuner, an extensible automatic tuning framework for programs. Chen et al. [\citealp{chen2018learning}] proposed a learning-based framework to optimize tensor programs. Cummins and others also introduced a portable, language-independent program graph representation [\citealp{cummins2021programl}], derived from compiler intermediate representations (IRs), as well as DeepTune [\citealp{cummins2017end}], a deep learning-based compiler optimization heuristic tool [\citealp{cummins2020deep}], demonstrating the potential of machine learning in automating tedious compiler tasks.


\subsection{Other Forms of Optimization}\label{2.3}

In other areas, research on efficiency optimization is extensive and profound. Krishnan et al. [\citealp{krishnan2018learning}] and Marcus et al. [\citealp{marcus2019neo}] studied methods for optimizing database system queries using deep learning. Schkufza et al. [\citealp{schkufza2013stochastic}] explored using random search to optimize x86 assembly code. These studies are particularly effective for programs that have no loops and are smaller in scale. However, our goal is to adopt generative methods to optimize the efficiency of higher-level source code.

\subsection{Code Understanding Datasets}\label{2.4}

Language modeling is an intriguing tool for code generation. For instance, the HumanEval dataset proposed by Chen et al. [\citealp{chen2021evaluating}] measures a model's programming capability through problem statements and function signatures. The APPS [\citealp{hendrycks2021measuring}] and CodeContests [\citealp{li2022competition}] datasets focus on assessing models' code generation abilities in complex programming contests. Meanwhile, domain-specific datasets such as DS-1000 [\citealp{lai2022ds}], GSM8K-Python [\citealp{chowdhery2022palm}], and MathQA-Python [\citealp{odena2021program}] are tailored for data science and mathematics respectively. Furthermore, datasets like PandasEval and NumpyEval [\citealp{zan2022cert}] are designed to evaluate code generation capabilities for specific libraries. Zan et al. [\citealp{zan2022language}] further introduced three proprietary library datasets: TorchDataEval, MonkeyEval, and BeatNumEval. In terms of code complexity prediction, Jeon et al. [\citealp{jeondeep}] introduced CodeComplex, the largest code dataset to date. The Code4Bench, developed by Majd et al. \mbox{[\citealp{majd2019code4bench}]}, is a comprehensive benchmark dataset widely used in program analysis techniques, covering 28 programming languages and containing a total of 3,421,357 programs, with 2,041,542 identified as having defects across 26 different categories. Unlike benchmarks such as Code4Bench, our research focuses on the efficiency optimization of Python code and proposes an innovative approach by creating efficient-inefficient code pairs through the concept of Algorithm Father-Son Pair, thereby optimizing code efficiency. Our goal is to enhance execution efficiency while maintaining code logic consistency, differentiating our method focused on algorithm logic and its execution efficiency optimization from the broad but relatively superficial coverage of Code4Bench.

\subsection{Code Cost Models}\label{2.5}

In the domain of code execution efficiency, cost models have been widely applied to predict code performance before runtime. Previous studies, such as the cost model for Halide programs by Adams et al. and the research on Tiramisu programs by Baghdadi et al., have demonstrated the effectiveness of cost models in performance prediction. Recently, the study by Zhou et al. \mbox{[\citealp{zhou2019deeptle}]}, introducing the DeepTLE model – a Bi-LSTM-based cost model – successfully predicted whether Python code would exceed the time limits set by the CodeForces website, saving up to 96\% in time cost. In contrast, our research employs a cost model based on LLMs, which not only predicts the likelihood of Python code exceeding time limits but also estimates the specific execution time of the code. The introduction of this novel approach provides a more precise and comprehensive perspective for predicting code performance.

Furthermore, Böck et al. \mbox{[\citealp{bock2023performance}]} utilized ASTNN (based on AST) \mbox{[\citealp{zhang2019novel}]}, to train a performance prediction model for C++ on the Code4Bench dataset. They specifically assessed the applicability of such models in the video game sector and found that transfer techniques (domain adversarial adaptation and model fine-tuning) were insufficient for transitioning these models to the targeted industrial domain of AAA games. Notably, they used incremental versions of problem solutions with different execution times from the same author's submissions, while we used efficient-inefficient code pairs. To contrast the differences in methodology, we trained six different cost models on the ACEOB-NPI dataset and analyzed the experimental results.

\subsection{Evaluating Large-Scale Language Models}\label{2.6}

Large Transformer architectures have achieved significant success in NL modeling, and their robust generative capabilities have been successfully applied to code generation. The CodeX system [\citealp{chen2021evaluating}] has demonstrated excellent performance and serves as the core of the advanced auto-completion system in GitHub Copilot. Austin et al. [\citealp{austin2021program}] also verified that refining a subset of the dataset can significantly increase the success rate of programming tasks. The AlphaCode system [\citealp{li2022competition}], based on the complete Transformers framework, proposed that efficiency can be improved through an unbalanced encoder-decoder ratio, accomplishing feats comparable to programmers.

ChatGPT [\citealp{chatgpt}] excels in a variety of tasks. ChatGPT-3.5 can understand context, provide relevant suggestions, and generate high-quality, syntactically correct code based on conversational inputs. Its subsequent version, ChatGPT-4, with a larger scale and more advanced architecture, has reached unprecedented proficiency in handling complex programming tasks by understanding more complex patterns and making accurate predictions, offering outstanding code generation performance.

Despite these advances, generating efficient code remains a significant challenge. Existing code generation systems still face the “slow positive” issue. Our ACEOB dataset aims to mitigate this issue by providing efficient-inefficient code pairs, promoting the generation of more efficient code.

\section{Human Coding Efficiency Survey}\label{3}

From the perspective of the current learning capabilities of AI, it can only glean “code patterns” from training datasets, yet it does not possess the understanding of how to compose efficient algorithms like a human programmer. Consequently, in order to enhance AI's learning ability, it is crucial to first discern what “code patterns” the existing AI has managed to acquire.

The study of human coding habits is pivotal in this respect, as LLMs are trained on code written by humans, thereby somewhat inheriting the characteristics of human coding. Thus, researching human coding habits equates to investigating the “code patterns” that LLMs have learned. This lays the foundation for developing an advanced understanding of AI capabilities and further informs the development of more efficient AI systems in the future.

We statistically analyzed the proportion of code execution time intervals for each level of difficulty (ranging from 0 to 27) in the ACEOB-Ori dataset. Specifically, we extracted 901,038 codes from 2,170,150 solution codes across 5,263 problems for this analysis. The rationale behind this extraction process is our observation that the problem distribution across different levels of difficulty on the website is markedly imbalanced. For instance, the lowest difficulty level (0) contained approximately 1,173,000 codes, which constitutes about 54\% of all the codes. This phenomenon suggests that programmers generally prefer simpler problems.

Our extraction method involved retrieving at most 500 codes for each problem, and uniformly extracting these at the same interval. For a problem of difficulty level $d$ (for example, $d$=0) and indexed at $i$, we scaled the execution time of all codes for that problem from 0\% to 100\%, and set six time breakpoints $TP(p,i,d)$. The calculation method of time breakpoints $TP(p,i,d)$ is as follows:
\begin{equation}
TP(p,i,d)= T_{min} ^{d-i}  + p \times (T_{max} ^{d-i} -T_{min} ^{d-i} ),p\in \left \{0 \% ,20 \% ,40 \%  ,60 \%  ,80 \% ,100 \% \right \} 
\end{equation}
where, $T_{min}^{d-i}$  and $T_{max}^{d-i}$  respectively represent the minimum and maximum execution time of all Python solution codes for the $i$ problem of difficulty $d$. To avoid the influence of extreme values, we excluded 1\% of the data. These six time breakpoints divide all solution codes into five parts according to their execution time. For instance, the fastest one-fifth codes of a problem would be categorized as “Fastest running time by 20\%”. For a time interval  $[TP(p,i,d), TP(p+20\%,i,d)]$, we calculated the number of codes $N_{[p, p+20\%]}^{d-i}$  for the $i$ problem of difficulty $d$ in this interval. Finally, for difficulty $d$, and interval  $[p, p+20\%]$  (for example, 0-20\%), the proportion $R([p, p+20\%],d)$ can be calculated as follows:
\begin{equation}
R([p, p+20 \%], d) = \frac{1}{n} \sum_{i=1}^{n} N_{[p, p+20\%]}^{d-i} 
\end{equation}
where, $n$ represents the number of problems at difficulty level $d$.

\begin{figure}[tb]
\begin{center}
\includegraphics[width=0.95\textwidth]{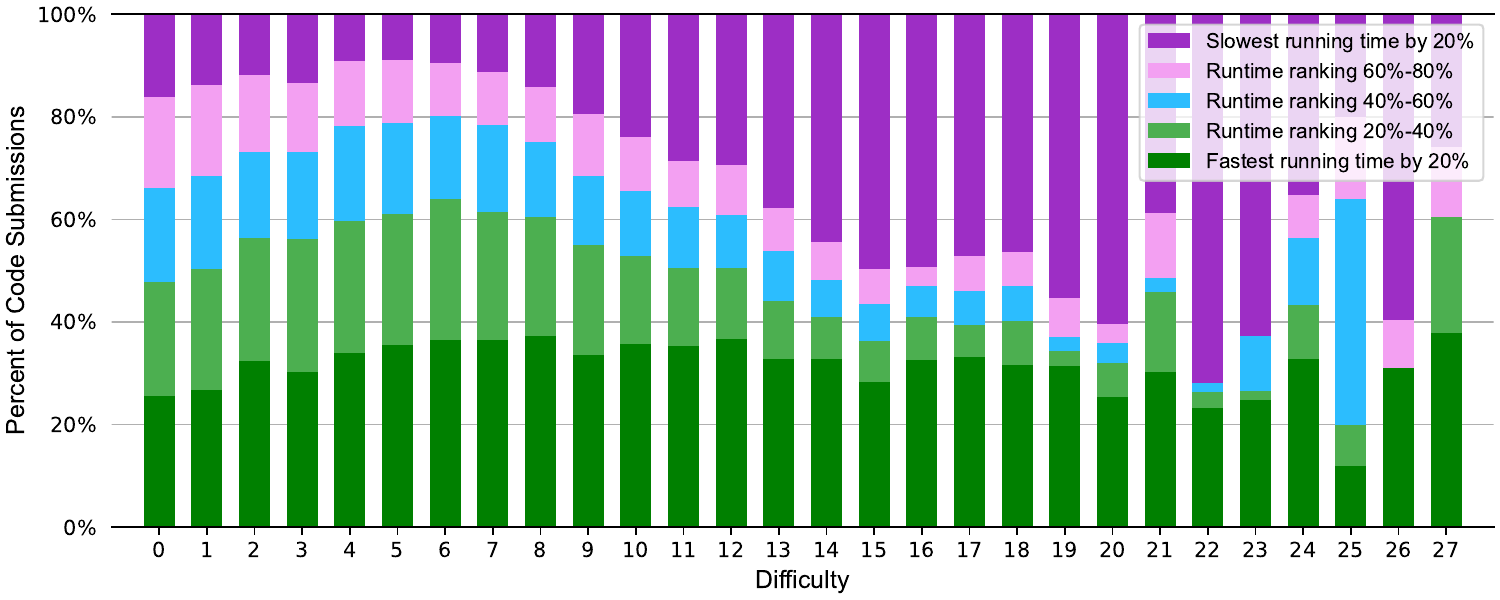}
\end{center}
\caption{The problem difficulty and the weighting of each code efficiency.}
\label{fig-difficulty-weighting}
\end{figure}

Our study has revealed a direct correlation between the quantity of inefficient code and problem difficulty. Figure \ref{fig-difficulty-weighting} illustrates the shifts in the proportion of code efficiency as the level of problem difficulty escalates. The research findings indicate that with the augmentation of problem complexity, the proportion of inefficient codes, classified as the “Slowest running time by 20\%”, continues to rise.

\begin{figure}[tb]
\begin{center}
\includegraphics[width=0.95\textwidth]{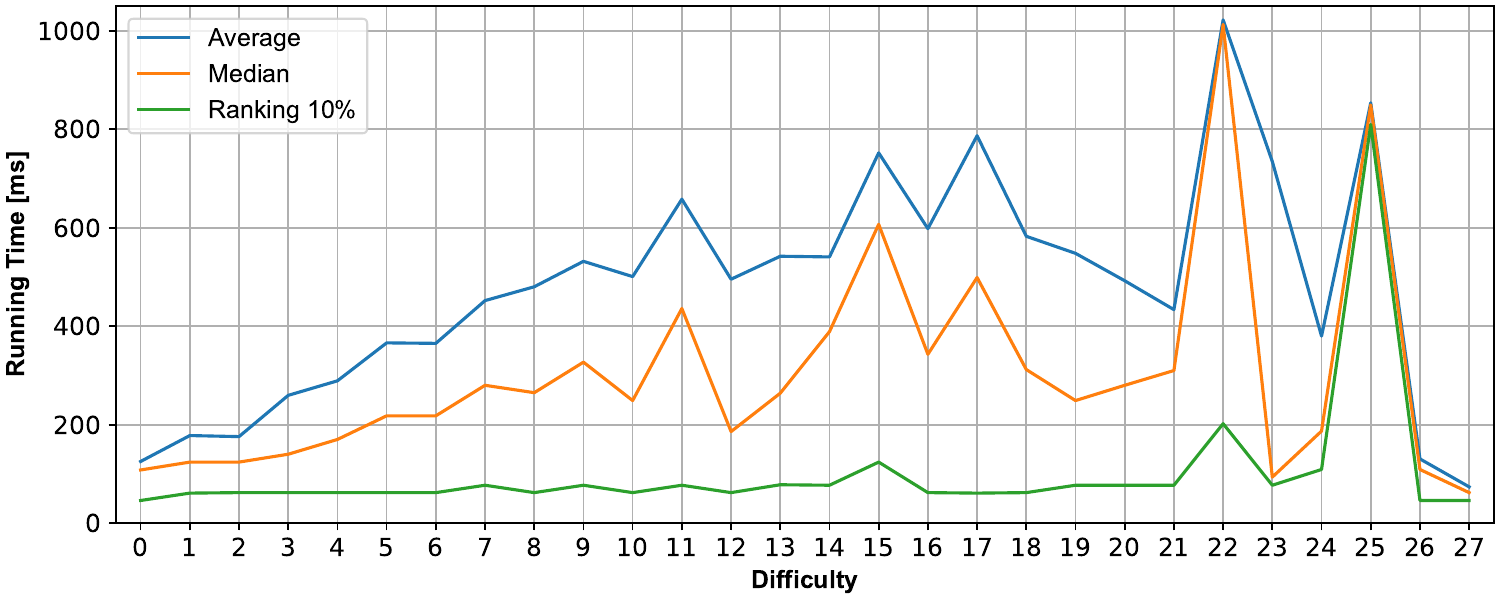}
\end{center}
\caption{The problem difficulty and the weighting of each code efficiency.}
\label{fig-line-difficulty}
\end{figure}

\begin{figure}[t]
\begin{center}
\includegraphics[width=0.95\textwidth]{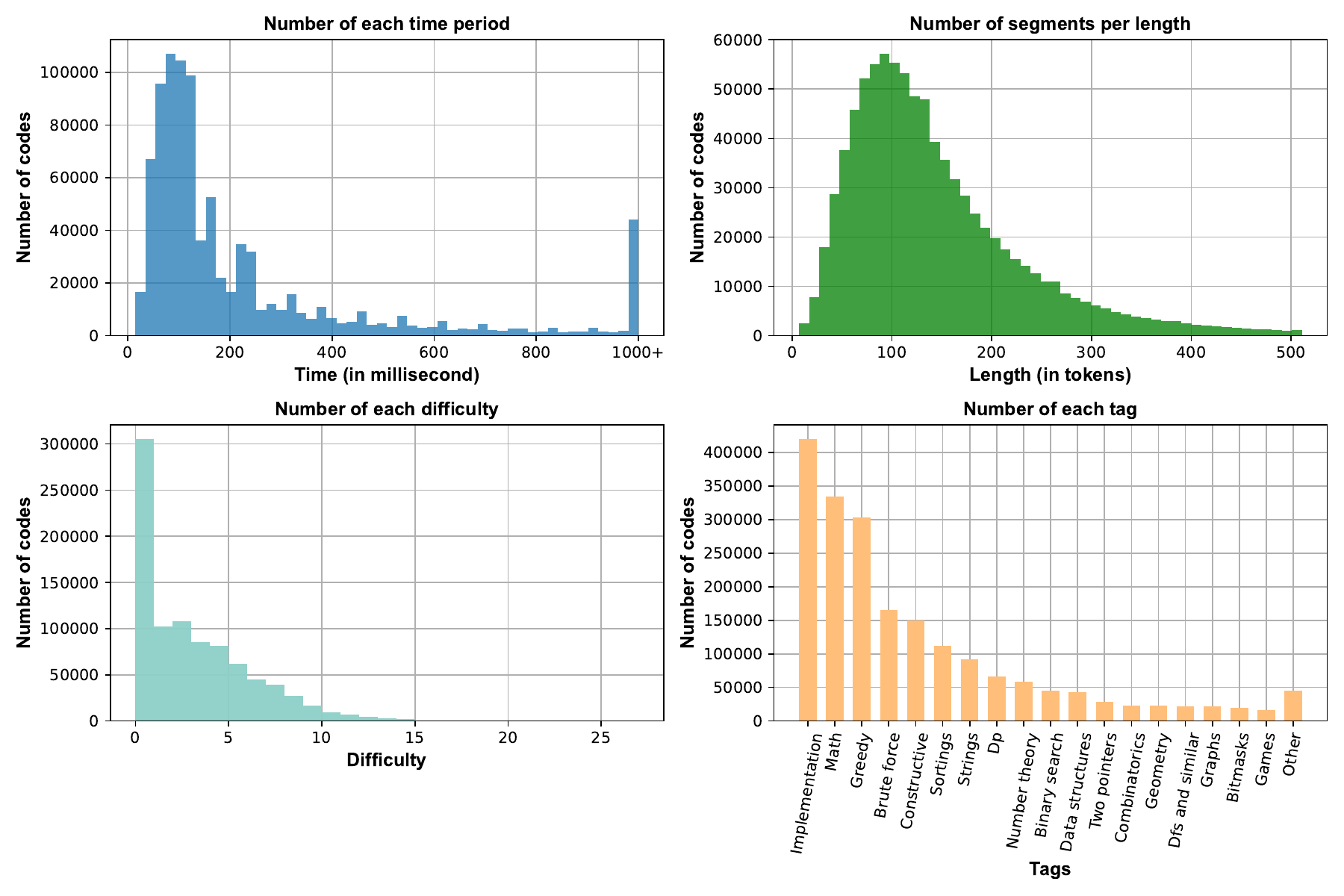}
\end{center}
\caption{Distribution proportions of the execution time of code (top left), token length (top right), problem-solving difficulty (bottom left), and algorithm tags (bottom right).}
\label{fig3}
\end{figure}


Further, we discerned that with increasing problem complexity, the disparity between the 10th percentile running time and the median continuously expands. Figure \ref{fig-line-difficulty} illustrates the relationship among the 10th percentile running time, the median, and the average, as the problem difficulty increases. The results demonstrate that the gap between the 10th percentile running time and the median broadens as problem complexity escalates, while the relationship between the median and the average remains nearly unchanged (due to a small number of problems in the difficulty range 23-27, accounting for only 0.38\% of all problems, the fluctuations become substantial).

Moreover, to delve into the proportions of different code lengths, execution times, difficulties, and algorithm tags in the dataset, we carried out further investigation. To prevent comments from affecting the results, we optimized the quality of all data. Figure \ref{fig3} suggests that most codes have short execution times, lengths centered around 100 tokens, difficulties focused on ultra-low levels, and primary algorithm tags are “implementation”, “math”, and “greedy”.

\section{What is IC2EC?}\label{4}

This section provides a detailed introduction to the definition of code efficiency optimization (\textbf{I}nefficient \textbf{C}ode to \textbf{E}fficient \textbf{C}ode, abbreviated as \textbf{IC2EC}) task and its input and output. The goal of the IC2EC task is to convert given \textbf{i}nefficient \textbf{c}ode (abbreviated as IC, indicating long running time) into functionally equivalent but more \textbf{e}fficient \textbf{c}ode (abbreviated as EC, indicating short running time). In this study, we focus on the capability of LLMs in handling the IC2EC task. Therefore, the task can be formalized as: EC = M(IC), where M represents the language model. In this modeling process, IC is considered as the “\textbf{feature}” input and EC as the “\textbf{label}” used for gradient propagation.

\begin{figure}[htbp]
\begin{center}
\includegraphics[width=0.95\textwidth]{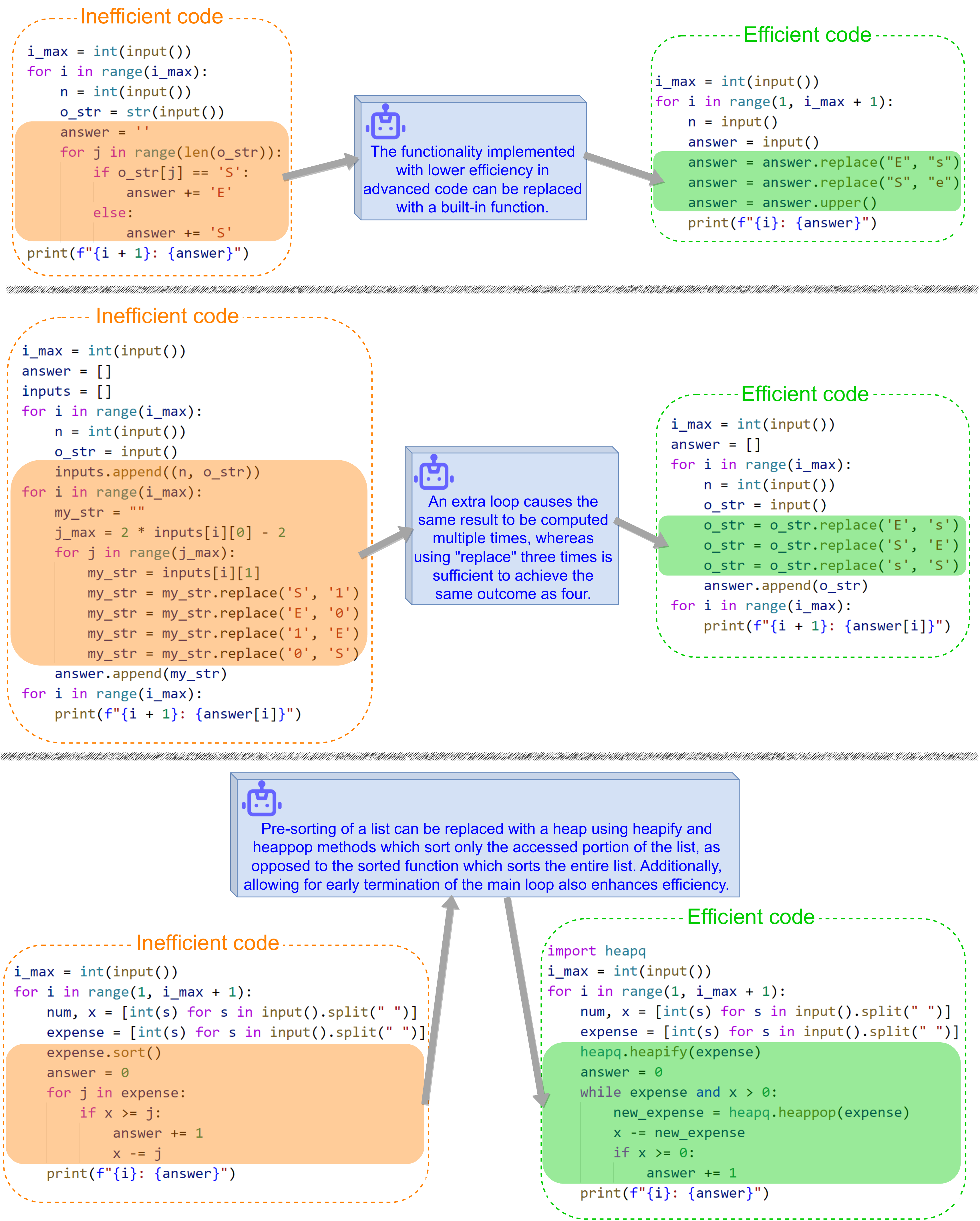}
\end{center}
\caption{Examples of the IC2EC task. Each line corresponds to an efficient-inefficient code pair, containing inefficient and efficient code. The three examples respectively emphasize the speed advantage of using built-in libraries (top), reducing the number of API calls (middle), and adopting more efficient data structures and algorithms (bottom).}
\label{fig-IC2EC-Examples}
\end{figure}

In analyzing the differences between IC and EC, we identified several viable optimization strategies. For instance, Figure \ref{fig-IC2EC-Examples} demonstrates the speed advantage of using built-in libraries, reducing the number of API calls, and adopting more efficient data structures and algorithms. These represent only a part of the various discrete design choices that LLMs can take to optimize code efficiency. We assume that LLMs can learn these discrete choices, thereby enhancing code efficiency through the transformation of IC into EC.

We define IC2EC as a \textbf{sequence-to-sequence problem,} that is, given an IC sequence, generate an EC version. Although sequence-to-sequence modeling is usually seen as a one-to-one problem, the IC2EC task is a one-to-many problem, as there can be multiple EC versions for a single IC. To reduce the interference caused by simple renaming of variables or syntax changes, we normalize the submitted code using common tokens when computing code similarity and scores.

In addressing the definition of IC2EC, we explored two core questions: \textbf{1)} Should the generated code focus more on efficiency or functionality? \textbf{2)} What are the limitations if the generated code requires manual correction?

Regarding \textbf{Question 1}, we believe that the generated code should focus more on \textbf{efficiency}. Since current LLMs often cannot generate code that is both fully functional and efficient, when faced with a choice, the IC2EC task should favor efficiency over functionality. This is because functionally deficient code can be perfected through manual modifications, which is easier than optimizing directly from IC. Meanwhile, code lacking in efficiency is redundant, as it is functionally identical and inefficient as the input IC. However, assessing the efficiency of code that cannot pass I/O unit tests is a challenge; this paper proposes the IOCCB and NPI metrics to address this issue.

Regarding \textbf{Question 2}, we believe that the factors affecting \textbf{manual correction} are the constraints. EC generated by LLMs that significantly deviates from the IC algorithmic approach may affect the speed of human understanding, and different variable naming styles may also impact the efficiency of code perfection. Therefore, to reduce the difficulty of manual modifications, we believe that the EC generated by LLMs should be as similar to the IC as possible. Our ACEOB dataset contains efficient-inefficient code pairs with similarity constraints to ensure efficiency optimization while improving the speed and accuracy of manual modifications.

In summary, although the definition of the IC2EC task is clear, enhancing the capabilities of LLMs to handle it still requires many meticulous techniques and strategies.

\section{Dataset Description and Usage}\label{5}
 
ACEOB is currently the first dataset specifically dedicated to competition-level Python code efficiency optimization. The training and test sets of this dataset are divided based on the temporal distribution of programming competition problems, with problems and their corresponding codes after \textbf{May 4th, 2022}, comprising the test set. ACEOB-NPI is a dataset used for training Python code cost models. Meanwhile, ACEOB-Ori provides a reference to the original dataset. Table \ref{tab-ACEOB-Data-items} details various data items and their descriptions within the ACEOB dataset. Compared to ACEOB, ACEOB-Ori does not differentiate between efficient-inefficient code pairs and alternate efficient codes, instead categorizing them uniformly as code data items. Furthermore, ACEOB-NPI only contains code data items. Tables \ref{tab-ACEOB-quantity} and \ref{tab-ACEOB-Ori-NPI} respectively provide the quantity description of various data items in the ACEOB, ACEOB-Ori, and ACEOB-NPI datasets.

\begin{table}[htbp]
\caption{Data items and descriptions within the ACEOB dataset.}
\centering
\begin{tabular}{| p{3.4cm} | p{12cm} |}

\hline
\textbf{Data Items}             &                \textbf{Description}  \\
\hline
Efficient-inefficient code pairs                                   &  Each entry in the ACEOB dataset contains a single pair of efficient-inefficient code pairs. These pairs are composed based on the concept of the Algorithm Father-Son Pair, with both codes having identical functionality but significantly different efficiencies. The names of these codes include execution time and memory information.   \\
\hline
Public IO unit tests                     &  These can be used as input and also provide preliminary screening functionality.  \\
\hline
Hide IO unit tests                       &  Used to evaluate the correctness of code functionality, i.e., whether it can complete the task.   \\
\hline
Alternate efficient codes                &  The collection of all efficient codes represents all the codes from the website that efficiently solve problems. In this paper, they are used to calculate IOCCB scores.   \\
\hline
Question dictionary                      &  Programming competition problems include problem descriptions in NL, input formats, output formats, and I/O test cases. We save various parts of the competition problems separately in the form of a dictionary.   \\
\hline
Difficulty                                &  The difficulty metric measures the complexity of the problem, which translates into the complexity of the code needed to implement its functionality. There are a total of 19 difficulty categories, ranging from the simplest introductory level 0 to the most challenging level 18 (sourced from the Codeforces website).   \\
\hline
Tags                                      &  A total of 36 algorithm tags. Algorithm tags denote the algorithmic approach used to solve the problem, such as brute force, geometry, and greed.   \\
\hline
Maximum time to achieve function          &  The longest execution time to implement the described functionality. This is derived from the longest execution time of the solution codes for the problem on the Codeforces website (excluding some extreme outlier data).   \\
\hline
Minimum time to achieve function          &  The shortest execution time to implement the described functionality. This is derived from the shortest execution time of the solution codes for the problem on the Codeforces website (excluding some extreme outlier data).   \\
\hline
Median time to achieve function           &  The median execution time to implement the described functionality.   \\
\hline
Problem Description Url                   &  The URL of the NL description of the problem on the Codeforces website.  \\
\hline
Solve Code Url                            &  The URL of the solution code for the problem on the Codeforces website.  \\
\hline
Title                                     &  The title of the problem. Due to the weak value of titles, we have not included them in the question dictionary.   \\
\hline
Time limit                                &  The time limit given for the problem on the Codeforces website.  \\
\hline
Memory limit                              &  The memory limit given for the problem on the Codeforces website.   \\
\hline

\end{tabular}
\label{tab-ACEOB-Data-items}
\end{table}

\begin{table}[thbp]
\caption{Quantity description of each data item in the ACEOB dataset.}
\centering
\begin{tabular}{| p{3.4cm} | p{12cm} |}

\hline
\textbf{Data Items}             &                \textbf{Quantity Description}  \\
\hline
Efficient-inefficient code pairs         &   A total of 95,359 data entries, of which 9,305 are designated as the test set. Each entry contains a pair of codes.          \\
\hline
Public IO unit tests                     &  On average, each entry contains 1.88 public I/O unit tests. \\
\hline
Hide IO unit tests                       &  On average, each entry contains 24.19 hidden I/O unit tests.   \\
\hline
Alternate efficient codes                &  On average, each entry contains 40.57 reference efficient codes.   \\
\hline
Question dictionary                      &  A total of 4,244 programming competition problems. The 95,359 entries are categorized into 4,244 problems, with 289 problems designated as the test set. These programming competition problems are difficult and complex, evidenced by the average length of their NL descriptions being 345 words.   \\
\hline
Difficulty                                &  19 difficulty categories. These are divided into three higher-level difficulty classifications: Introductory (level 0), Interview (levels 1-3), and Competition (levels 4-18). In the ACEOB test set of 9,305 entries, these three difficulty classifications have 5,372 (Introductory), 2,178 (Interview), and 1,755 (Competition) entries respectively.   \\
\hline
Tags                                      &  A total of 36 algorithm tags. On average, each entry contains 2.26 algorithm tags. Figure \ref{fig-algorithm-tags} shows the distribution of algorithm tags in the 9,305 entries of the ACEOB test set.   \\
\hline

\end{tabular}
\label{tab-ACEOB-quantity}
\end{table}

\begin{figure}[thbp]
\begin{center}
\includegraphics[width=0.95\textwidth]{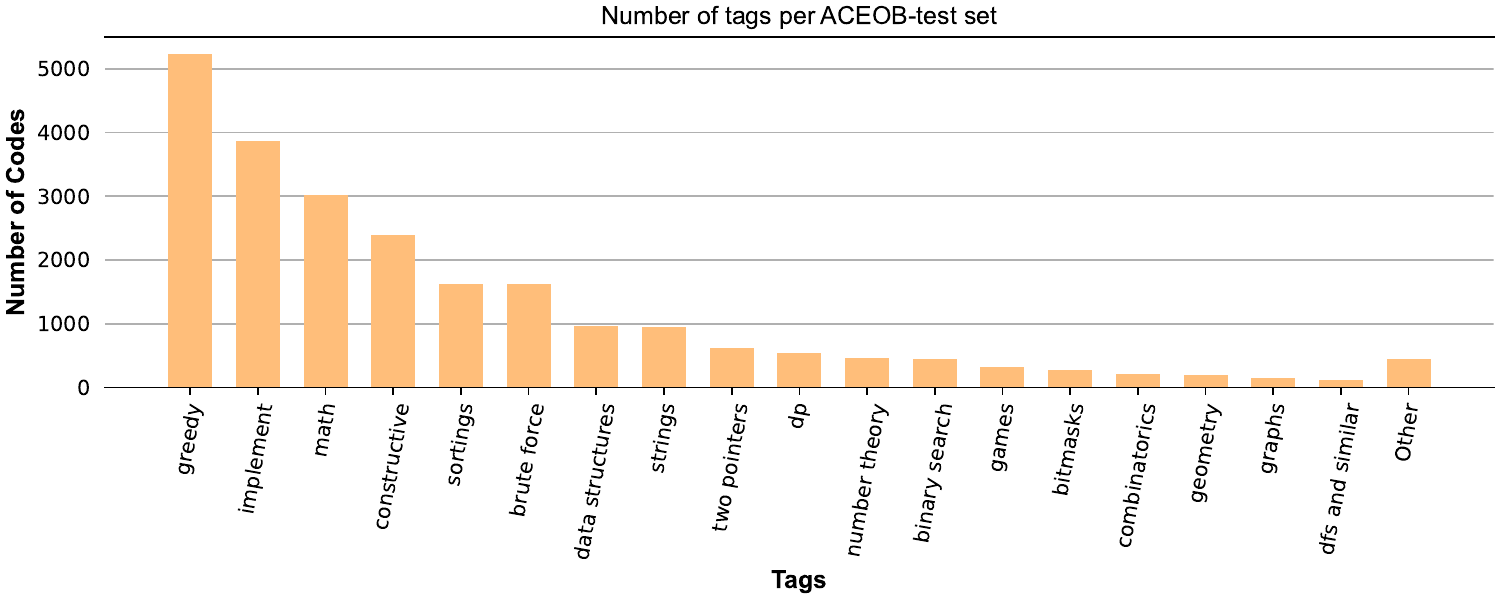}
\end{center}
\caption{The distribution of algorithm tags is presented within the 9,415 samples of the test set derived from the ACEOB dataset.}
\label{fig-algorithm-tags}
\end{figure}

\begin{table}[thbp]
\caption{Quantity description of each data item in the ACEOB-Ori (top) and ACEOB-NPI (bottom) datasets.}
\centering
\begin{tabular}{| p{1.8cm} | p{2.6cm} | p{10.5cm} |}

\hline
\textbf{Dataset}                         &   \textbf{Data Items}                      &                \textbf{Quantity Description}  \\
\hline
\multirow{5}{*}{ACEOB-Ori}   
&  Codes                                    &   A total of 897,941 data entries. Each entry contains a code.          \\
\cline{2-3}
&  Public IO unit tests                    & On average, each entry contains 1.84 public I/O unit tests. \\
\cline{2-3}
&  Hide IO unit tests                       & On average, each entry contains 23.69 hidden I/O unit tests.  \\
\cline{2-3}
&  Question dictionary                      &  A total of 5,262 programming competition problems. The 897,941 entries are categorized into 5,262 problems. The average length of the NL descriptions of the programming competition problems is 356 words. \\
\cline{2-3}
&  Difficulty                                &  28 difficulty categories. Ranging from the simplest introductory level 0 to the most challenging level 27 (sourced from the Codeforces website).  \\
\cline{2-3}
&  Tags                                      &   A total of 36 algorithm tags. On average, each entry contains 2.21 algorithm tags.\\
\hline
ACEOB-NPI &  Codes                                 &  A total of 617,691 code data entries, of which 12,354 are designated as the test set. The code names include execution time information and NPI scores.   \\
\hline

\end{tabular}
\label{tab-ACEOB-Ori-NPI}
\end{table}

\begin{figure}[tbhp]
\begin{center}
\includegraphics[width=0.8\textwidth]{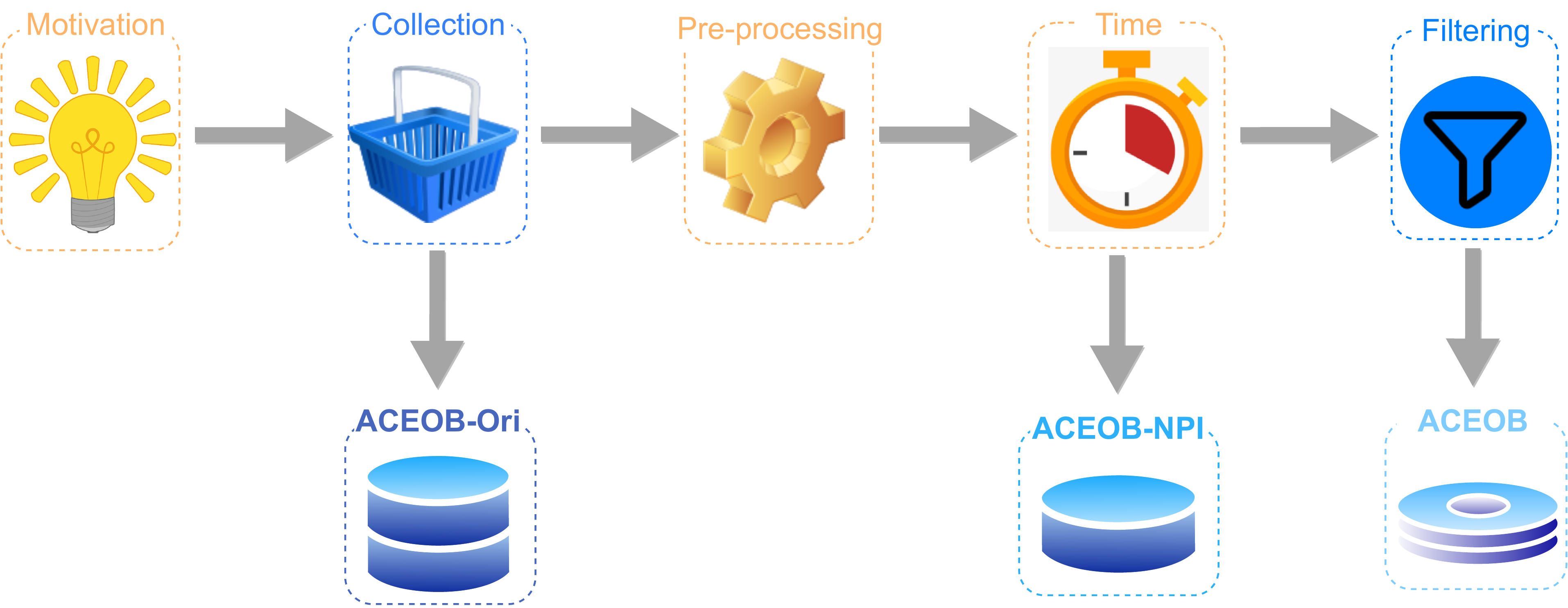}
\end{center}
\caption{Process of dataset creation from scratch.}
\label{fig-Dataset-creation-process}
\end{figure}

The ACEOB dataset is an ideal choice for training LLMs for the IC2EC task in Python code. Each data item in the dataset contains a pair of efficient-inefficient code, which are formed based on our proposed \textbf{Algorithm Father-Son Pair} concept (see Section \mbox{\ref{7.2}}). ACEOB is suitable for Python code efficiency optimization scenarios where IC is used as input, and it also supports scenarios that require more information input, as it includes various information such as programming competition problems. In terms of extended applications, ACEOB, with its included programming competition problems, solution codes, and I/O tests, can be used for code generation. Due to containing different codes implementing the same function, ACEOB can also be extended for use in code semantic cloning research.

The ACEOB-Ori and ACEOB-NPI datasets are each suitable for different research tasks. ACEOB-Ori provides original code data, supporting a broader range of dataset research and applications, while ACEOB-NPI is specifically suitable for training cost models for Python code execution times. Additionally, as ACEOB-NPI includes the NPI score information of codes, it can also be used to train cost models that predict Python code NPI scores, thereby enhancing the precision and application scope of code efficiency analysis.

\section{Dataset Construction}\label{6}

This section introduces the construction process and respective roles of three datasets: ACEOB-Ori, ACEOB-NPI, and ACEOB. Figure \ref{fig-Dataset-creation-process} illustrates the entire construction process flowchart. Initially, we collected a vast array of code data from the Codeforces programming contest website to build the ACEOB-Ori original dataset. Subsequently, based on ACEOB-Ori, we executed a series of data preprocessing steps to enhance code quality, including code normalization and AST purification. Then, we re-evaluated and recorded the execution time of each code segment to create the ACEOB-NPI dataset, focusing on code efficiency prediction. Finally, by implementing the Algorithm Father-Son Pair method, we streamlined the number of data items, resulting in the final ACEOB dataset.

\subsection{Motivation for Dataset Construction}\label{6.1}

Elevating the capabilities of LLMs in the domain of code efficiency optimization is a challenging task. The crux lies in establishing an appropriate benchmark dataset. This dataset should contain pairs of code that are functionally equivalent but differ in efficiency. Although existing datasets like CodeContests [\citealp{li2022competition}] are extensive, they lack a comparative focus on efficiency. To meet the specific needs of the IC2EC task, we built the ACEOB dataset from scratch.

\subsection{Data Collection}\label{6.2}

We crawled a vast amount of programming contest code from the \mbox{\href{https://codeforces.com}{Codeforces} website, committed to creating a high-quality dataset.} Codeforces is one of the most popular and well-known coding challenge and practice platforms worldwide. To construct a high-quality dataset, we set strict data filtering criteria, requiring us to have a large source of data. The number of solutions under different programming contest problems varies significantly. For problems with a higher number of solutions, we ranked them based on execution time, crawled at fixed intervals, and set an upper limit (500 codes). For problems where the code quantity did not meet this limit, we conducted a complete crawl of all solutions.

A quality dataset should encompass exhaustive data information. To create a dataset that meets the requirements of the IC2EC task, we collected hidden I/O unit tests, problem difficulty, and tags, among other information. These are key to verifying the correctness of code functionality and optimizing code efficiency. To enhance the dataset's versatility and practicality, we collected not only the code itself but also the natural language description of the problem, input/output format, I/O unit test samples, time/space constraints, and the source URLs of the problem and code (Table \ref{tab-ACEOB-Data-items}). We developed a custom HTML parser to correctly format LaTeX expressions in the problem text. For example, the symbol \textbf{$\sum $} would be represented as \textbf{\textbackslash sum}. Additionally, we also recorded each piece of code's space usage. Although the IC2EC task may not directly utilize this additional information, it is included in the dataset to promote and support related field research.

Through this series of collection and processing, we eventually formed the ACEOB-Ori dataset and presented a quantitative description of each type of data item in Table \ref{tab-ACEOB-Ori-NPI} (top). ACEOB-Ori contains all the information shown in Table \ref{tab-ACEOB-Data-items}, laying the foundation for further data processing and analysis.

\subsection{Data Pre-processing}\label{6.3}

To enhance the dataset's quality, we performed a series of data pre-processing steps on the code data directly collected from the website. This is because the original data often varies in quality, including issues such as overly long codes, numerous meaningless comments, and special code commands (e.g., setting recursion limits).

\begin{itemize}
\item \textbf{Basic Normalization:} We removed special code operations and comments. Since Codeforces is primarily used for competitions, many codes are written for execution rather than reading. The comments are often annotations on erroneous fragments and are considered noise in the IC2EC task.
\item \textbf{AST Purification:} We converted the code into AST and then back into code to address inappropriate line breaks and compilation issues. This step significantly improved the quality of code data.
\item \textbf{Deletion of Duplicate Code:} We utilized the Longest Continuous Common Substring algorithm to calculate code similarity and deleted codes with a similarity greater than 0.99.
\item \textbf{Deletion of Overly Long Code:}  We set relative and absolute length limits and removed overly long codes from each problem. Relative length limits consider the average length of all codes for each problem, while absolute length limits are based on the model's maximum input limitation, removing codes exceeding 512 tokens (the maximum input length for some models).
\end{itemize}

\subsection{Re-evaluate Dataset Code Execution Time}\label{6.4}

We reevaluated the execution time of the codes in the dataset. Due to significant errors in the estimated execution time for the same code on Codeforces (up to 200 ms), we adopted a standard practice in performance engineering: executing each code 30 times and using the median as the execution time to minimize the impact of minor variations. Additionally, we standardized the execution time, considering the differences in the number of hidden I/O tests on Codeforces. We converted the code's execution time to correspond to the execution time of 47 I/O unit tests. Moreover, we removed some codes where verifying the correctness of code functionality was extremely challenging.

\subsection{ACEOB-NPI: Python Code Cost Model Training Set}\label{6.5}

Current research on Python code cost models is insufficient. In the domain of compiler automatic optimization, while there exist small neural network models predicting program runtime, such as the execution time prediction models for Halide and Tiramisu code proposed by Adams [\citealp{adams2019learning}] and Baghdadi [\citealp{baghdadi2021deep}], these are primarily used for low-level algorithm acceleration and are not applicable to Python code. With the advancement of LLMs, cost model predictions based on LLMs are expected to achieve better results.

To this end, we created the ACEOB-NPI dataset. We calculated the NPI score for each code in the dataset and added it as information to the code names. Additionally, we divided the dataset into training and test sets at a 50:1 ratio, constituting the ACEOB-NPI dataset. Since each piece of code data in ACEOB-NPI contains execution time and NPI score information, it can be used to train cost models predicting the execution time of Python codes as well as predicting the NPI scores of Python codes. This paper proposes two cost models based on LLMs, aimed at predicting execution time and NPI scores respectively, to meet different research needs.

\subsection{Dataset Similarity Reduction}\label{6.6}

Low similarity among codes in a dataset is a key indicator of its high quality. While Hendrycks et al. [\citealp{hendrycks2021measuring}] used \textbf{T}erm \textbf{F}requency\textbf{-}\textbf{I}nverse \textbf{D}ocument \textbf{F}requency (\textbf{TF-IDF}) features combined with \textbf{S}ingular \textbf{V}alue \textbf{D}ecomposition (\textbf{SVD}) dimension reduction and cosine similarity for deduplication in the APPS dataset, we believe there are superior alternative methods.

We found that TF-IDF features are not suitable for code data. TF-IDF is a statistical measure primarily used to evaluate the relevance of a word in a collection of documents. This is applicable to NL text but not to code, as the criticality in code is often reflected in the logical structure rather than individual variable naming [\citealp{lachaux2021dobf}]. Therefore, we did not adopt TF-IDF features for removing duplicate code data.

Instead, we adopted \textbf{CodeBLEU} as the metric for assessing code similarity. CodeBLEU combines the strengths of BLEU [\citealp{papineni2002bleu}] in n-gram matching and incorporates code syntax and semantics considerations through AST and data flow. Consequently, CodeBLEU can more accurately assess code similarity and serve as a similarity measure for clustering algorithms.

We employed clustering methods to reduce redundancy. To adapt to the characteristics of code data, we used CodeBLEU scores as the similarity measure for clustering algorithms, instead of traditional measures like Euclidean distance. To avoid discrepancies in CodeBLEU scores due to different variable or function names, we standardized variable and function names before clustering (e.g., variables \textbf{a} and \textbf{b} would be standardized to \textbf{var1} and \textbf{var2}) and restored them afterwards. This approach focuses more on the substantive similarity of the code rather than superficial naming differences.

Using a modified Bisecting k-means algorithm [\citealp{arthur2007k}], we independently clustered sets of efficient and inefficient codes, with the number of clusters being proportional to the square of the total number of codes. We selected the code with the highest sum of CodeBLEU scores from other codes in each cluster as the representative of the cluster. Following the Algorithm Father-Son Pair concept (see Section \ref{7.2}), these representative codes were grouped into efficient-inefficient code pairs. Through this method, we filtered down the 897,941 codes of ACEOB-Ori to approximately 22\%, forming the ACEOB dataset.

\subsection{ACEOB Dataset}\label{6.7}

Each data item in the ACEOB dataset is an efficient-inefficient code pair, constructed according to the Algorithm Father-Son Pair concept (see Section \ref{7.2}). However, this pairing method assumes that each IC corresponds to a unique EC, which is not consistent with reality as typically an IC might have multiple efficient solutions. Therefore, we added alternate efficient codes to each efficient-inefficient code pair, which are efficient codes using different algorithms. These alternate efficient codes are used to calculate the IOCCB scores, enriching the dataset's utility and effectiveness.

To avoid data leakage, we adopted a temporal split method to divide the training and test sets in the dataset. We allocated problems and their codes after May 4, 2022 (the release date of the last model, PolyCoder, in the experimental section) to the test set, ensuring the independence and validity of the test set. This method is superior to the random splitting approach used in APPS, which risks including test set data in the model's pretraining dataset. Our temporal split method ensures the novelty and challenge of the test set, aligning more closely with real-world application scenarios.

\section{Approach}\label{7}

\subsection{IC2EC Task System}\label{7.1}

\begin{figure}[t]
\begin{center}
\includegraphics[width=0.8\textwidth]{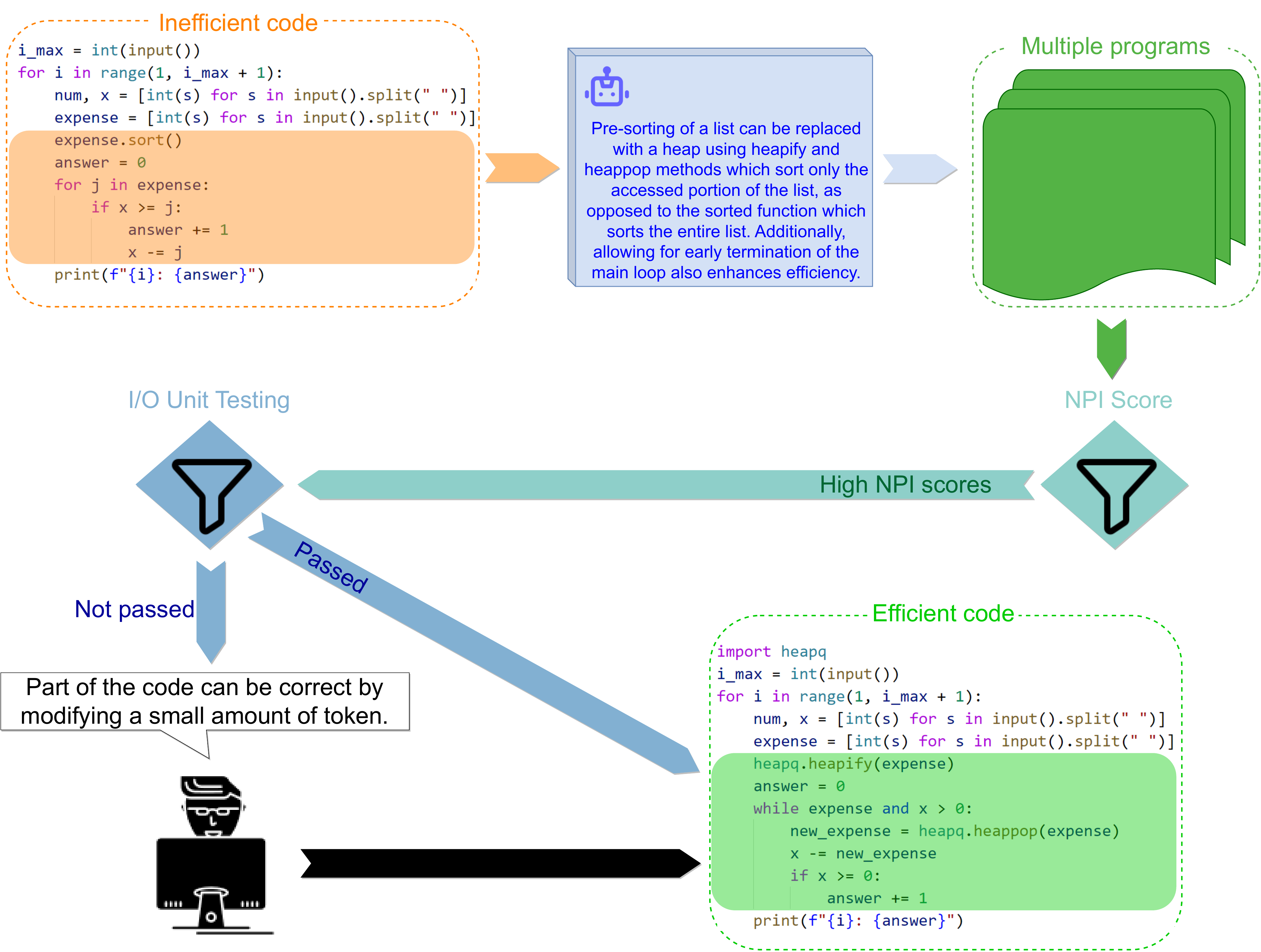}
\end{center}
\caption{High-level View of the IC2EC Task System. This figure presents a high-level view of the code efficiency optimization system. The system first utilizes LLMs to receive IC and generates multiple preliminarily optimized code versions. Subsequently, these generated code versions are filtered through an NPI filter, selecting the more efficient code. Finally, code that passes the I/O unit testing is directly submitted, while code that fails the test is further repaired by professional programmers.}
\label{fig-IC2EC-System}
\end{figure}

We have designed a system aimed at automatically generating EC based on a given IC. Figure \ref{fig-IC2EC-System} provides an overview of the system for the IC to EC task. Our system design and evaluation are based on the following two core principles:
\begin{itemize}
\item \textbf{Efficiency First:}  The generated code must be an improvement in terms of efficiency. We allow for the generated code to have some functional defects, as we believe these can be fine-tuned and fixed manually; therefore, the primary consideration for the model's output is the improvement in efficiency.
\item \textbf{Consideration of the One-to-Many Problem:}  For a given IC, there may exist multiple discrete transformations to generate more efficient code. For instance, altering data structures, optimizing loop boundaries, or improving function encapsulation. Since these potential edits are unknown a priori, the model needs to automatically learn which transformations can enhance efficiency and generate code that is logically closer to the original IC, thereby reducing the cost for developers to understand the new code.
\end{itemize}

\subsection{Selection of Ground-Truth Efficient Code}\label{7.2}

An important consideration during model training is the selection of \textbf{ground-truth} for gradient propagation. We recognize that for a given IC, there are multiple possible EC, and which one is chosen for gradient propagation will directly impact the training effectiveness. We advocate selecting the EC that, while ensuring its efficiency and functionality, is most algorithmically similar to the IC. As current LLMs generated code often requires programmer review, this can reduce the time cost of manual understanding.

We introduce the concept of Algorithm Father-Son Pair to guide the selection process of ground-truth EC. This concept selects pairs of IC and EC with the same algorithm logic as an Algorithm Father-Son Pair. In the ACEOB dataset, we use the CodeBLEU score as a standard for algorithmic similarity. Under the premise of unifying variable and function names, we select the code from EC that has the highest CodeBLEU score in comparison to IC, with the requirement that they perform identical functions. We then employ the Hungarian Algorithm [\citealp{kuhn1955hungarian}] to form efficient-inefficient code pairs as training data.

\subsection{IOCCB Metric}\label{7.3}

\begin{figure}[thbp]
\begin{center}
\includegraphics[width=0.9\textwidth]{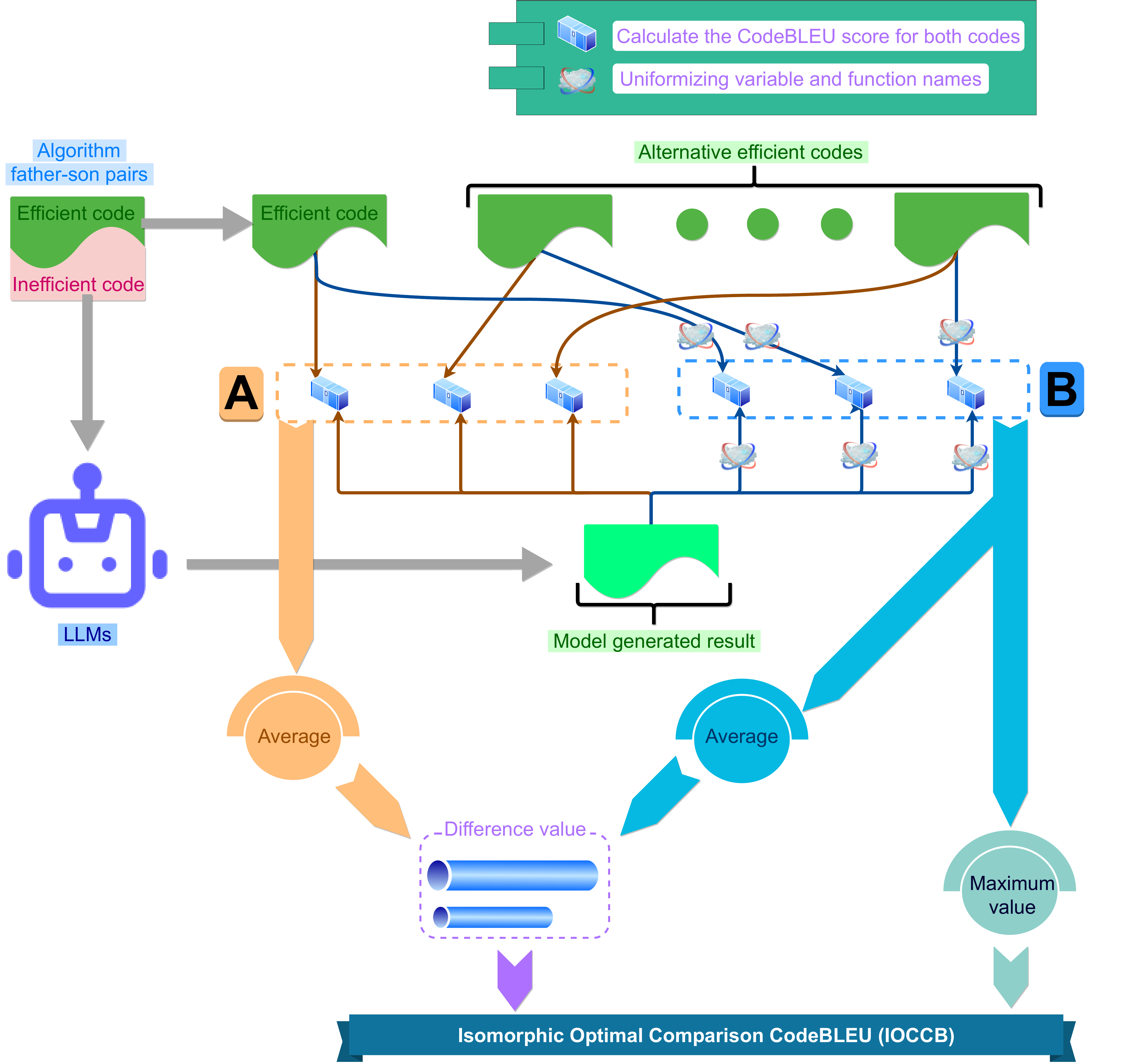}
\end{center}
\caption{IOCCB Score Calculation Process. This figure shows the detailed calculation process of the IOCCB score. The process begins with inputting the IC from the Algorithm Father-Son Pair into LLMs to generate the predicted code $g$. Then, by matching the generated code $g$ with ECs and alternate efficient codes to calculate the CodeBLEU score, forming the set $O$. Additionally, each code will be standardized in terms of variables and function names before matching, forming the set $S$. Next, we calculate the average of set $O$ ($O_{avg}$), the average of set $S$ ($S_{avg}$), and the maximum of set 
$S$ ($S_{max}$). Finally, the IOCCB score is defined as the maximum of set $S$ plus the square root of the difference between the average of set $S$ and set $O$, i.e., $B_{max}+\sqrt{(B_{avg}-A_{avg})}$.}
\label{fig-IOCCB-Calculation}
\end{figure}

Despite CodeBLEU [\citealp{ren2020codebleu}] leveraging the advantages of the BLEU [\citealp{papineni2002bleu}] metric and enhancing the consideration for code syntax and semantics through AST and data flow, it exhibits significant score variation when dealing with codes that have different variable naming, failing to reflect the essential similarity between codes. Moreover, due to multiple ECs corresponding to a single IC, the CodeBLEU score of a single EC may not comprehensively evaluate the efficiency of the generated code.

Therefore, we propose the \textbf{I}somorphic \textbf{O}ptimal \textbf{C}omparison \textbf{C}ode\textbf{B}LEU (\textbf{IOCCB}) metric, aiming to more accurately measure the efficiency and similarity of the generated code by standardizing variable and function names and comparing them with multiple answers to select the optimal matching value. Figure \ref{fig-IOCCB-Calculation} illustrates the calculation process of the IOCCB score. The features of the IOCCB metric include:
\begin{itemize}
\item \textbf{Maximizing Scores Across Multiple ECs:} As the IC2EC task is a one-to-many problem, a low CodeBLEU score for model-generated code compared to the ground-truth EC does not necessarily imply poor code quality, as it may adopt a different algorithm and be more similar to other EC. Therefore, we calculate the CodeBLEU score of the model-generated result against all ECs in alternate efficient codes and take the maximum value as the evaluation criterion.
\item \textbf{Standardization of Variable and Function Names:}  Considering that differences in variable and function naming should not affect the assessment of code functionality and efficiency, we standardize all codes when calculating the IOCCB score. To compensate for the potential change in CodeBLEU score distribution due to standardization, we also calculate the average difference of CodeBLEU scores before and after standardization and adjust the score accordingly to ensure the accuracy and fairness of the IOCCB score.
\end{itemize}

\subsection{NPI Metric}\label{7.4}

Traditional metrics such as time complexity, execution time, and acceleration have a long history in evaluating code efficiency, but they have limitations when assessing the ability of LLMs to perform IC2EC tasks. Especially when the functions implemented by the code are different, using execution time or acceleration to judge the efficiency of the code is insufficient. For example, if a code implementing a simple function takes 100 ms and another implementing a complex function takes 500 ms, we cannot simply assume the former is more efficient than the latter. Time complexity can evaluate the efficiency of an algorithm, but its determinacy in worst-case scenarios is unsolvable [\citealp{turing1938computable}]. Moreover, execution time and acceleration alone cannot accurately measure the potential optimization space of code after model optimization.

\begin{figure}[b]
\begin{center}
\includegraphics[width=0.8\textwidth]{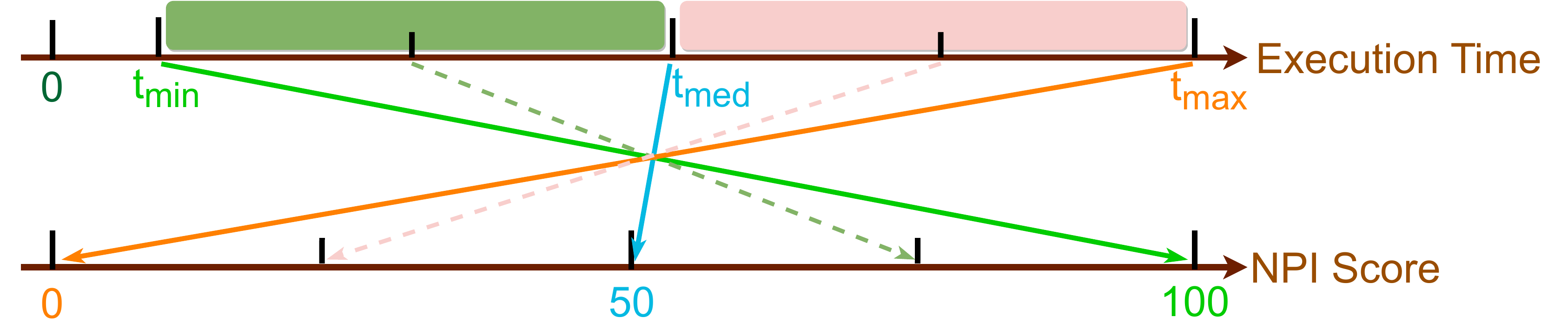}
\end{center}
\caption{NPI Metric Calculation Process. This figure details the calculation process of the NPI metric, which is implemented in two steps. First, the execution time's median and minimum values are mapped to the scoring range [50, 100]. Then, the execution time's maximum value and median are mapped to the scoring range [0, 50]. The three key points, 0, 50, and 100, represent the maximum execution time, median execution time, and minimum execution time, respectively.}
\label{fig-NPI-Calculation}
\end{figure}

In response to these challenges, we propose the \textbf{N}ormalized \textbf{P}erformance \textbf{I}ndex (\textbf{NPI}) metric, an indicator to evaluate the capability of LLMs in the IC2EC task. As the average execution time for codes implementing different functions varies, NPI emphasizes reflecting the efficiency range and optimization potential of the code by standardizing execution time within the 0 to 100 range. Figure \mbox{\ref{fig-NPI-Calculation}} illustrates the NPI metrics calculation process. For a given problem $p$ and its solution code $c$, the $NPI_{c}$ score is calculated as follows:
\begin{equation}
NPI_{c}  = 
\begin{cases}
(\frac{ t_{med} - t_{c}} {t_{med}  - t_{min}} ) \times 50 + 50
  & \text{ if } t_{c} <  t_{med}
\\
               50 
  & \text{ if } t_{c} =  t_{med} 
\\
(\frac{ t_{max} - t_{c}} {t_{max}  - t_{med}} ) \times 50 
  & \text{ if } t_{c} >  t_{med}
\end{cases}
\label{formula4}
\end{equation}
where $t_{c}$ represents the execution time of code $c$. Correspondingly, $t_{max}$, $t_{med}$, and $t_{min}$ denote the maximum, median, and minimum execution times among all known code samples, respectively. It is pertinent to note that the known code samples for problem $p$, as specified in our study, are sourced from all submissions for this problem on the Codeforces website.

The NPI metric provides an accurate assessment of the LLMs' capability to perform IC2EC tasks by reflecting the current code's efficiency position and optimization space relative to other codes. The NPI score can reflect the efficiency range of the current code and the extent to which the code can be optimized. NPI is also capable of measuring the efficiency of codes with different functions and individual isolated codes.

\subsection{Python Code Cost Model \& NPI Filter}\label{7.5}

Although research on Python code cost models is relatively scarce, they play an important role in the IC2EC task. When the model optimizes IC, its output must be efficient even if there are functional deficiencies (which may only require simple manual modifications). Otherwise, even if the code can pass I/O testing, its inefficiency makes it merely equivalent to another IC, almost without value. Therefore, the value premise of the output results is that they must be efficient, which requires cost models to assess.

We have trained two major cost models: predicting Python code execution time and predicting Python code NPI scores. Since predicting NPI is more challenging, the cost model accuracy for predicting execution time is higher than that for predicting NPI. However, the NPI score can reflect more information (such as remaining optimization space information) and provide effective guidance, which may be more valuable than execution time in some cases. Therefore, we provide two types of cost models.

The NPI filter is a mechanism that uses NPI scores to filter inefficient codes. In experiments, the NPI filter first uses the cost model that predicts execution time to calculate the code's execution time, then calculates the code's NPI score using this execution time. Based on the NPI score, the filter will select efficient codes, providing an effective preliminary filtering method for optimizing inefficient codes.

\section{Experiment}\label{8}

In this section, we present the experimental design and its outcomes, aiming to address the following pivotal research questions:
\begin{itemize}
\item \textbf{RQ1:} What is the fit of IOCCB and NPI metrics to code efficiency? How do the two Python code cost models compare in terms of fit?
\item \textbf{RQ2:} In the IC2EC task, how do decoder models differ in performance from encoder-decoder models? What is the optimization effect of our NPI filter on both types of models?
\item \textbf{RQ3:}  In the IC2EC task, what is the performance comparison between beam search and sampling? What are their respective areas of strength? After introducing the NPI filter, how do these two methods compare in terms of performance optimization?
\item \textbf{RQ4:} How do various models perform as the difficulty level of programming competition problems increases?
\item \textbf{RQ5:} How do various models perform when dealing with problems with different algorithmic labels?
\item \textbf{RQ6:} How does the ChatGPT model perform in the IC2EC task and other efficiency-related domains?
\end{itemize}

\subsection{Models}\label{8.1}

We conducted fine-tuning evaluations on the ACEOB dataset with various CodeLMs, including CodeGen [\citealp{nijkamp2022codegen}], PolyCode [\citealp{xu2022systematic}], and CodeT5 series models [\citealp{wang2021codet5}]. Both CodeGen and PolyCode are decoder models, while the CodeT5 series is based on the T5 model as encoder-decoder models, specifically for code generation. Additionally, we evaluated the performance of the ChatGPT model [\citealp{chatgpt}] on the ACEOB dataset. The specific models are as follows:
\begin{itemize}
\item \textbf{CodeT5-small (60M).} As the smallest scale model, CodeT5-small offers good performance with lower resource requirements.
\item \textbf{CodeT5-base (220M).} A medium-scale CodeT5 model that balances performance and resource consumption.
\item \textbf{CodeT5-large-ntp-py (770M).} The largest scale CodeT5 model, further pretrained on an additional Python dataset, focusing on Python code generation. Note that this model originates from Le et al.'s CodeRL research [\citealp{le2022coderl}].
\item \textbf{CodeGen-mono (350M).} A GPT-2 based CodeGen model, supporting multiple programming languages and demonstrating good performance.
\item \textbf{PolyCoder (0.4B).} The PolyCode model is a deep learning model focusing on encoding and decoding multiple programming languages, supporting code autogeneration and program understanding.
\item \textbf{ChatGPT.} ChatGPT possesses strong capabilities in code generation. It not only understands and interprets programming requirements but also generates corresponding code snippets based on these requirements, effectively enhancing development efficiency. Especially in dealing with common programming tasks and problems, its prediction and code generation capabilities are remarkably accurate. Here, we employ ChatGPT-3.5.
\end{itemize}

\subsection{Fine-tuning Configuration}\label{8.2}

In the fine-tuning process on the ACEOB dataset, our goal was to optimize given IC into more EC. We employed both beam search and sampling methods for comparison. All LLMs, including cost models, were fine-tuned using the AdamW optimizer and a learning rate scheduler with a linear decay preheating phase. Training settings included 10 epochs, a batch size of 32, a maximum length of 512 tokens, and sampling parameters set to top-k=50, top-p=0.95, and temperature=0.8. Other parameters followed the default settings of Hugging Face [\citealp{huggingface}].

We fine-tuned two types of Python code cost models, both based on CodeT5-base, on the ACEOB-NPI dataset. Unlike the IC2EC task, the cost models' input “\textbf{feature}” is the code, while the “\textbf{label}” is either execution time or NPI score.

To guide the generation of specific content, we designed particular prompts for fine-tuning LLMs and cost models:
\begin{itemize}
\item \textbf{Fine-tuning LLMs model:} “Please optimize the following inefficient code into a more efficient version, while keeping the functionality unchanged:\textbackslash n \{inefficient code\}\textbackslash n More efficient version:\textbackslash n”.
\item \textbf{Fine-tuning cost models:} “\{code\}\textbackslash n Evaluate the code efficiency score: ”.
\end{itemize}

\subsection{Comprehensive Comparison of Model Efficiency Optimization Strategies}\label{8.3}

We conducted a comprehensive comparison of nine different configurations adopted by various models to assess their performance in IC2EC tasks. In these settings, NPI refers to the submission of code with the highest NPI score through an NPI filter. The specific configurations are as follows:
\begin{itemize}
\item \textbf{G:}  reedy search is used, selecting the token with the highest probability at each step.
\item \textbf{B5:}   Implements beam search with a beam width of 5, generating five pieces of code and submitting the first one, i.e., the code with the highest probability.
\item \textbf{B5+NPI:}  Similar to B5, but submits the code with the highest NPI score through the NPI filter.
\item \textbf{S5:}  Generates five pieces of code using sampling and randomly submits one.
\item \textbf{S5+NPI:}  Similar to S5, but submits the code with the highest NPI score.
\item \textbf{B10:}  Implements beam search with a beam width of 10, generating ten pieces of code and submitting the first one, i.e., the code with the highest probability.
\item \textbf{B10+NPI:}  Similar to B10, but submits the code with the highest NPI score through the NPI filter.
\item \textbf{S10:}  Generates ten pieces of code using sampling and randomly submits one.
\item \textbf{S10+NPI:}  Similar to S10, but submits the code with the highest NPI score.
\end{itemize}

In these configurations, we utilized IOCCB and NPI scores to measure the efficiency of the code. IOCCB score focuses on evaluating the token similarity of the code, while the NPI score concentrates on measuring the relative efficiency of the optimized code. Additionally, we employed I/O Pass Rate (Pass@1) as a metric to measure the functionality of the code. This comprehensive comparison aims to deeply understand the performance of different optimization strategies, as well as the extent of the impact of the NPI filter on the efficiency of the generated code. Through this approach, we hope to uncover the areas of strength and potential directions for improvement for various configurations, providing an empirical foundation for future research on IC2EC tasks.

\subsection{RQ1: Fit Assessment of Evaluation Metrics \& Cost Models}\label{8.4}

\begin{table}[tbhp]
\caption{Analysis of the Spearman rank correlation coefficients for the IOCCB and NPI evaluation metrics in the ChatGPT generated code set and the ACEOB test set, focusing on execution time and the prediction accuracy of two cost models: predicted NPI and execution time.}
\centering
\begin{tabular*}{0.9\textwidth}{@{\extracolsep{\fill}}c c c c c}
\hline
                    &   \multicolumn{2}{c}{\textbf{ChatGPT Generation Set}}                           &   \multicolumn{2}{c}{\textbf{ACEOB Test Set}}  \\
\cline{2-5} 

\textbf{Models}    
&    {\textbf{\textit{Correlation}}}    &    {\textbf{\textit{P-value}}} 
&    {\textbf{\textit{Correlation}}}    &    {\textbf{\textit{P-value}}} 
\\
\hline
CodeBLEU \textbf{\&} Run Time                         &  -0.1488                     &  $e^{-29}$                    &  -0.6252               &  $e^{-324}$  \\
IOCCB \textbf{\&} Run Time                            &  -0.2166                     &  $e^{-60}$                    &  -0.5393      &  $e^{-324}$  \\
\hline
Predicted NPI \textbf{\&} True NPI                    &  \textbf{0.3931}             &  $e^{-207}$                   &  0.604                 &  $e^{-324}$   \\
Predicted Run Time \textbf{\&} True Run Time          &  0.2185                      &  $e^{-61}$                    &  \textbf{0.7099}                &  $e^{-324}$   \\
\hline
\end{tabular*}
\label{tab-Spearman-fit}
\end{table}

\begin{figure}[t]
\begin{center}
\includegraphics[width=0.95\textwidth]{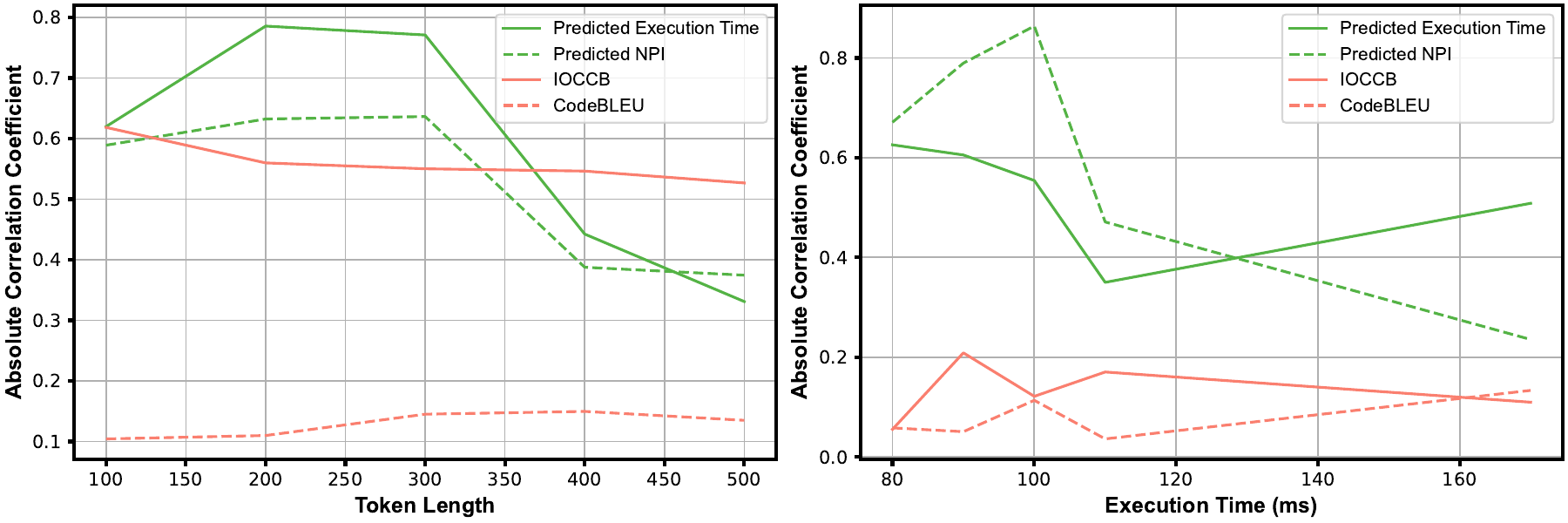}
\end{center}
\caption{The impact of code length (left) and execution time (right) on the absolute values of the correlation coefficients of CodeBLEU, IOCCB, predicted NPI, and predicted execution time across four indicators.}
\label{fig-Spearman-length-time}
\end{figure}

We utilized the Spearman Rank Correlation Coefficient [\citealp{spearman1961proof}], a non-parametric statistical tool measuring the rank-order relationship between two variables, to assess the fitting of IOCCB and NPI metrics to code efficiency. This coefficient ranges from -1 to 1, where 1 indicates a perfect positive correlation and -1 indicates a perfect negative correlation. Notably, this coefficient is robust to outliers and doesn't require the data to be normally distributed or linearly related, making it particularly suitable for this study.

We selected a set of human-written codes (ACEOB Test Set) and codes generated by LLMs (ChatGPT Generation Set) as our evaluation datasets. Considering that the ChatGPT generated set is not entirely correct, we only considered the part that passed the IO tests (59.65\%).

The IOCCB score is a specially designed variant of CodeBLEU for code efficiency. Table \ref{tab-Spearman-fit} (top) shows the Spearman rank correlation coefficient of the two metrics with execution time. We compared the Spearman Rank Correlation of CodeBLEU and IOCCB scores with execution time and found that the correlation coefficient of the IOCCB score with execution time is significantly higher than that of the CodeBLEU score, particularly in the set of human-written codes. This finding validates the superiority of the IOCCB score in measuring code efficiency.

The NPI score is a standardized rating system that maps code execution time to a scoring range, highlighting the efficiency of code execution. To verify the predictive accuracy of the NPI cost models for code NPI scores, we compared the Spearman Rank Correlation between predicted NPI scores (predicted by the predictive NPI cost model) and actual NPI scores (calculated through Eq. \mbox{\ref{formula4}} and actual execution time). Table \mbox{\ref{tab-Spearman-fit}} (bottom) presents the Spearman rank correlation coefficient of predicted and actual values of NPI and execution time for both cost models. The results indicate that both cost models have a high correlation with actual NPI scores and a very small p-value, indicating significant statistical correlation (typically, we consider there to be significant statistical correlation if the p-value is less than 0.05).

When assessing the code set generated by ChatGPT, all indicators had lower correlation than with the real dataset. This may be due to significant characteristic differences between ChatGPT-generated code and human-written code, and the generated dataset containing only optimized single codes rather than efficient-inefficient code pairs.

\begin{table}[bt]
\caption{RMSE results of six Python cost models on the ACEOB-NPI dataset, with ZeroR as the baseline, predicting the mean of the training set.}
\centering
\begin{tabular*}{0.8\textwidth}{@{\extracolsep{\fill}} l l | c c}

\hline
 &  & \multicolumn{2}{c}{\textbf{RMSE (ms)}} \\ 

\cline{3-4} 

         \textbf{Model}         &       \textbf{Model Input}             & \textbf{Predicted NPI}    & \textbf{Predicted Execution Time}   \\
                  
\hline

Null (ZeroR)         &  Null (ZeroR)            & 29.65               & 204.16              \\ 
\hline
CodeT5              & Split AST              & 38.72	           &  201.31          \\ 
\hline
ASTNN               & Split AST              &  22.32           &	151.14             \\ 
\hline
\textbf{CodeT5}              & \textbf{Code}                  & \textbf{21.36}               & \textbf{105.49}              \\ 
\hline

\end{tabular*}
\label{tab-ASTNN-RMSE}
\end{table}

Building on the work of Böck et al. \mbox{[\citealp{bock2023performance}]}, who applied the ASTNN method \mbox{[\citealp{zhang2019novel}]} to train a cost model for C++, this paper further conducts comparative experiments in training Python cost models using the ASTNN approach. As shown in Table \mbox{\ref{tab-ASTNN-RMSE}}, on the ACEOB-NPI dataset, this study fine-tuned and compared six Python cost models. The model prediction methods are categorized into two types: predicting NPI and predicting execution time; the baseline models include ASTNN and CodeT5; the model inputs are divided into source code and segmented AST (adopting the input method of the ASTNN approach). Additionally, a Zero Rule Estimator (ZeroR) was introduced as a baseline, which predicts the mean of the training set. For regression tasks, the results on the test set are reported in the form of Root Mean Square Error (RMSE), calculated as follows:
\begin{equation}
RMSE = \sqrt{\frac{1}{n} \sum_{1}^{n}(y_i - \hat{y}_i)^2} 
\label{formula5}
\end{equation}
where \mbox{$y_i$} represents the actual cost for code \mbox{$i$}, and \mbox{$\hat{y}_i$} is the predicted cost.

Table \mbox{\ref{tab-ASTNN-RMSE}} displays the RMSE performance of six Python cost models on the ACEOB-NPI dataset. We found that CodeT5 and ASTNN demonstrated better performance. Particularly, the ASTNN model, when dealing with segmented AST data, showed near-optimal performance, especially in predicting NPI. Notably, when the CodeT5 model directly takes the source code as input, its prediction accuracy reached the highest, highlighting the significant advantage of CodeT5 in handling raw code as opposed to dealing with segmented AST data. Consequently, we decided to adopt the configuration of the CodeT5 model with source code input as the optimal setup for the evaluation metrics model in this study.

We conducted a more in-depth analysis for RQ1. Figure \mbox{\ref{fig-Spearman-length-time}} shows the impact of code token length and execution time on the absolute values of the correlation coefficients of CodeBLEU, IOCCB, and the two cost model indicators.

In Figure \ref{fig-Spearman-length-time} (left), as the token length increases, the correlation coefficient for predicted execution time first increases then decreases, peaking in the 250-300 token length range; the predicted NPI shows a similar trend but peaks slightly earlier than predicted execution time. The correlation coefficient of IOCCB decreases with increasing token length, while that of CodeBLEU slightly increases.

In Figure \mbox{\ref{fig-Spearman-length-time}} (right), as execution time increases, the correlation coefficient for predicted execution time first decreases then increases, with the valley appearing at 110 ms; the correlation coefficient for predicted NPI shows the opposite trend, peaking at around 100 ms and then rapidly decreasing. IOCCB slightly decreases, while CodeBLEU's correlation slightly increases with increasing execution time.

\begin{center}
\begin{tcolorbox}[colback=gray!10,
                  colframe=black,
                  width=16.2cm,
                  arc=1mm, auto outer arc, 
                  boxrule=0.5pt, %
                  toprule=1.2pt, rightrule=1.2pt, bottomrule=1.2pt,leftrule=1.2pt, 
                  ]

In summary, we conclude the following:
\begin{itemize}
\item The IOCCB metric significantly outperforms the CodeBLEU metric.
\item The CodeT5 model, when utilizing source code as input, outperforms models processing AST-segmented data.
\item The IOCCB metric is inversely proportional to code token length and execution time, while the CodeBLEU metric is directly proportional to these variables.
\item Both cost models have their advantages, with the predicted execution time model being more precise, while the predicted NPI model provides more informative insights.
\item Regarding code length, both cost models perform best in the 250-300 token length range; regarding execution time, the predicted execution time model and predicted NPI model show opposite trends, with the former first decreasing then increasing and the latter first increasing then decreasing.
\end{itemize}
\end{tcolorbox}
\end{center}



\subsection{RQ2: NPI Filter and Model Performance Evaluation}\label{8.5}

\begin{figure}[t]
\begin{center}
\includegraphics[width=0.95\textwidth]{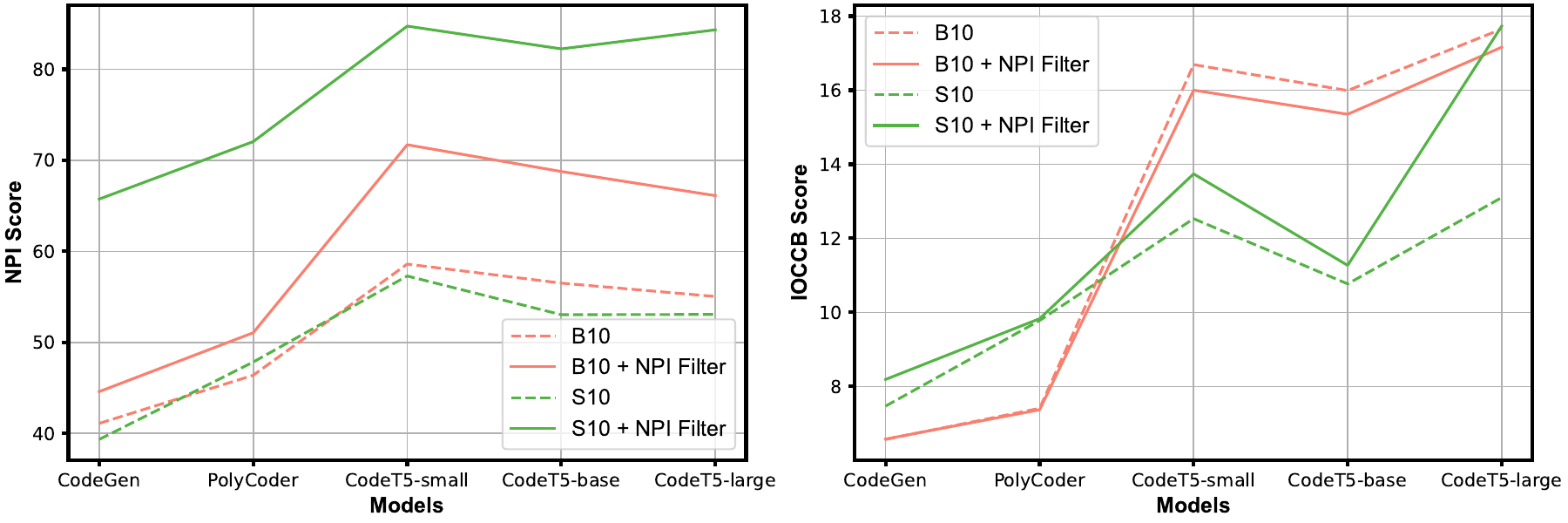}
\end{center}
\caption{The impact of the NPI filter on the performance of different models using beam and sampling methods.}
\label{fig_B10_S10_NPI}
\end{figure}

From the experimental data in Tables \ref{tab-Overall-decoder}, \ref{tab-Overall-en-decoder}, and Figure \ref{fig_B10_S10_NPI}, we observed several key trends.

\begin{table}[thbp]
\caption{Overall performance of decoder models CodeGen and PolyCoder on the ACEOB test set.}
\begin{center}
\begin{tabularx}{0.9\textwidth}{X X X r}
\hline
\textbf{Models}     & \textbf{\textit{I/O}} (\%)   &    \textbf{\textit{NPI}}   &   \textbf{\textit{IOCCB}} \\
\hline 
CodeGen(G)  & 0   &   39.51   &   7.38   \\
\hline
B5       & 0   &   38.77   &   6.68   \\
B5+NPI  & 0   &   41.11   &   6.69   \\
S5        & 0   &   39.44   &   7.43   \\
S5+NPI  & 0   &   59.2   &   7.88   \\
\hline
B10       & 0   &   41.1   &   6.57   \\
B10+NPI  & 0   &   44.59   &   6.58   \\
S10       & 0   &   39.34   &   7.47   \\
S10+NPI  & 0   &   65.74   &   8.19  \\
\hline
\rowcolor{gray!15}  PolyCoder(G)  & 0   &   47.98   &   \textbf{9.87}   \\
\hline
\rowcolor{gray!15}  B5       & 0   &   45.06   &   7.37   \\
\rowcolor{gray!15}  B5+NPI  & 0   &   48.05   &   7.34   \\
\rowcolor{gray!15}  S5        & 0   &   47.46   &   9.77   \\
\rowcolor{gray!15}  S5+NPI  & 0   &   65.88   &   9.83   \\
\hline
\rowcolor{gray!15}  B10       & 0   &   46.39   &   7.41   \\
\rowcolor{gray!15}  B10+NPI  & 0   &   51.06   &   7.37   \\
\rowcolor{gray!15}  S10       & 0   &   47.84   &   9.78   \\
\rowcolor{gray!15}  S10+NPI  & 0   &   \textbf{72.05}   &   9.83  \\
\hline

\textbf{Average}  &  0  &  \textbf{48.92}  &  \textbf{7.97} \\
\hline
\end{tabularx}
\label{tab-Overall-decoder}
\end{center}
\end{table}


\begin{table}[thbp]
\caption{Overall performance of encoder-decoder models CodeT5 and ChatGPT on the ACEOB test set.}
\begin{center}
\begin{tabularx}{0.9\textwidth}{X X X r}
\hline
\textbf{Models}     & \textbf{\textit{I/O}} (\%)   &    \textbf{\textit{NPI}}   &   \textbf{\textit{IOCCB}} \\
\hline
CodeT5-small(G)  & 0.03   &   60.07   &   12.94   \\
\hline
B5       & 0.15   &   58.83   &   14.42   \\
B5+NPI  & 0.13   &   68.82   &   14.26   \\
S5        & 0.06   &   58.04   &   12.55   \\
S5+NPI  & 0.05   &   79.6   &   13.5   \\
\hline
B10       & 0.46   &   58.6   &   16.69   \\
B10+NPI  & 0.19   &   71.71   &   16.0   \\
S10       & 0   &   57.29   &   12.53   \\
S10+NPI  & 0.05   &   \textbf{84.74}   &   13.74  \\
\hline
\rowcolor{gray!15}  CodeT5-base(G)  & 0.13   &   54.55   &   10.91   \\
\hline
\rowcolor{gray!15}  B5       & 0.42   &   55.51   &   12.87   \\
\rowcolor{gray!15}  B5+NPI  & 0.32   &   64.19   &   12.72   \\
\rowcolor{gray!15}  S5        & 0.13   &   53.29   &   10.81   \\
\rowcolor{gray!15}  S5+NPI  & 0.17   &   76.36   &   11.16   \\
\hline
\rowcolor{gray!15}  B10       & 0.85   &   56.51   &   15.99   \\
\rowcolor{gray!15}  B10+NPI  & 0.55   &   68.77   &   15.35   \\
\rowcolor{gray!15}  S10       & 0.13   &   53.04   &   10.77   \\
\rowcolor{gray!15}  S10+NPI  & 0.12   &   82.24   &   11.27  \\
\hline
CodeT5-large(G)  & 0.3   &   53.24   &   11.94   \\
\hline
B5       & 0.78   &   52.83   &   15.86   \\
B5+NPI  & 0.67   &   60.93   &   15.38   \\
S5        & 0.26   &   53.27   &   12.84   \\
S5+NPI  & 0.24   &   77.76   &   15.82   \\
\hline
B10       & 0.95   &   55.04   &   17.65   \\
B10+NPI  & 0.7   &   66.12   &   17.16   \\
S10       & 0.27   &   53.07   &   13.1   \\
S10+NPI  & 0.2   &   84.33   &   17.73  \\
\hline
\hline 
\rowcolor{gray!15}  \textbf{Average}    &  \textbf{0.31}  &   \textbf{63.66}  &   \textbf{13.92}  \\
\hline
\textbf{ChatGPT(G)}  &     \textbf{59.69}   &   54.88   &   \textbf{30.8}       \\
\hline
\end{tabularx}
\label{tab-Overall-en-decoder}
\end{center}
\end{table}


Firstly, the application of the NPI filter produced significant effects. Across all models and settings, configurations using the NPI filter (i.e., submitting the code with the highest NPI score) generally outperformed those without the filter. For instance, in the PolyCoder model, the S10 configuration without the filter had an NPI score of 47.84, while the B5+NPI configuration with the NPI filter increased the score to 72.05. Additionally, we found that random submission post-sampling is typically the worst-performing configuration out of the nine. However, when combined with the NPI filter, configurations that sample ten codes and apply the NPI filter generally perform best. This strongly indicates the effectiveness of the NPI score and suggests that our NPI filter can be transferred to other models as a code quality filtration tool.

Secondly, for I/O metrics, most models scored 0, with only the CodeT5 series and ChatGPT showing slight fluctuations. This indicates that current Large Language Models still have limitations in understanding code syntax and implementing code functionality. However, while ChatGPT performs well on I/O metrics, it does not excel in other metrics, suggesting that its learning in code efficiency is not as advanced as its learning in code syntax.

\begin{figure}[thbp]
\begin{center}
\includegraphics[width=0.95\textwidth]{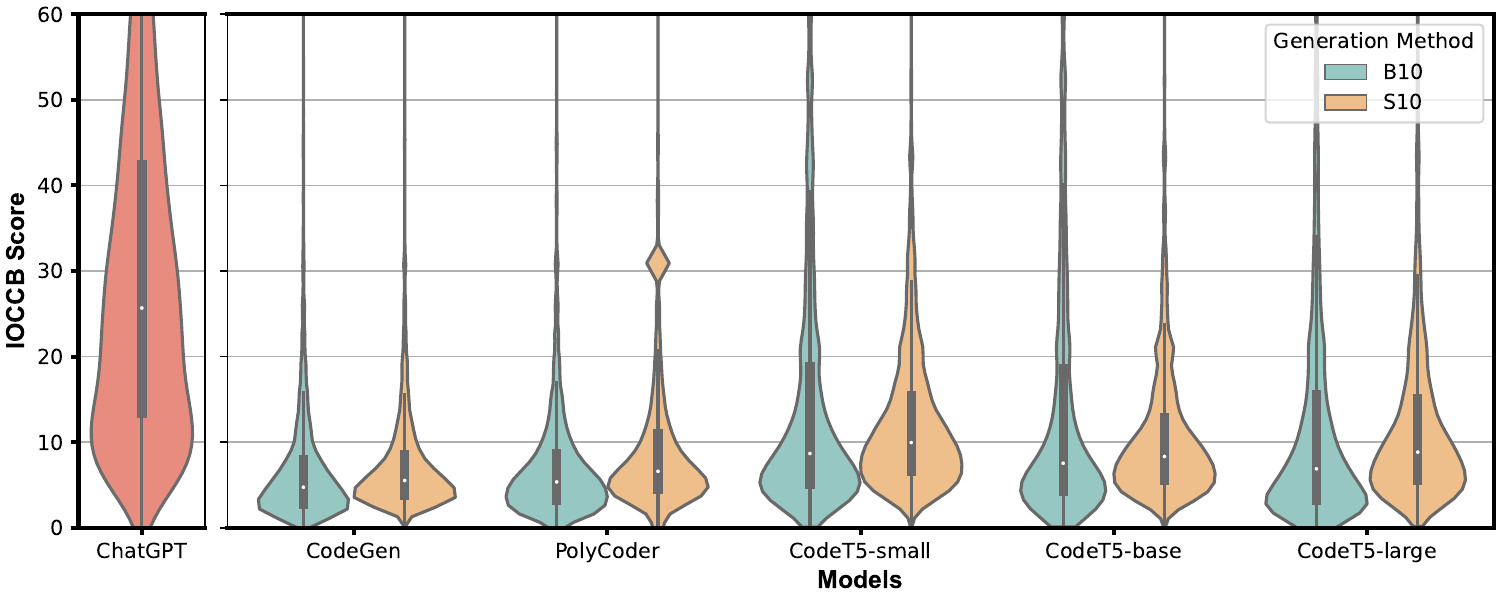}
\end{center}
\caption{Distribution of IOCCB scores for code generated by different models.}
\label{fig-IOCCB-Violin}
\end{figure}

Regarding the IOCCB metric, our NPI filter was able to slightly improve model performance in most cases. Figure \ref{fig-IOCCB-Violin} clearly shows the distribution of IOCCB scores for different model-generated codes, indicating that encoder-decoder models significantly outperform decoder models.

\begin{figure}[thbp]
\begin{center}
\includegraphics[width=0.95\textwidth]{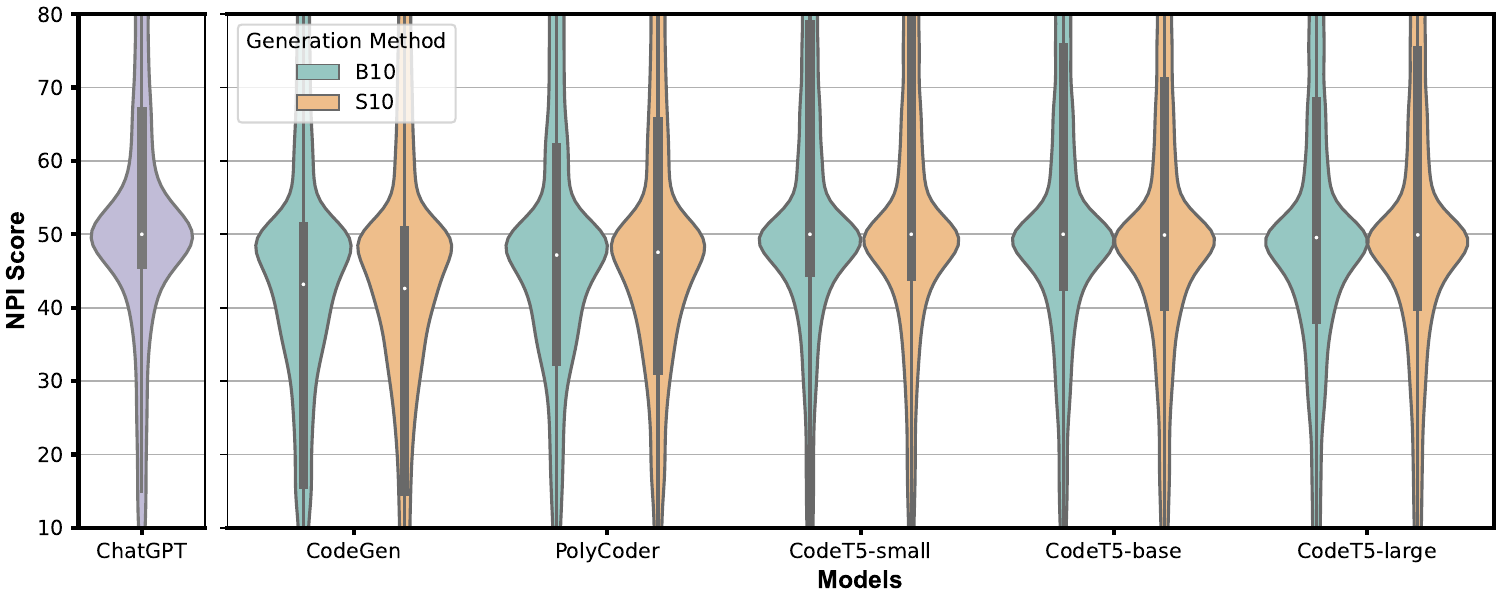}
\end{center}
\caption{Distribution of NPI scores for code generated by different models.}
\label{fig-NPI-Violin}
\end{figure}

\begin{figure}[thbp]
\begin{center}
\includegraphics[width=0.95\textwidth]{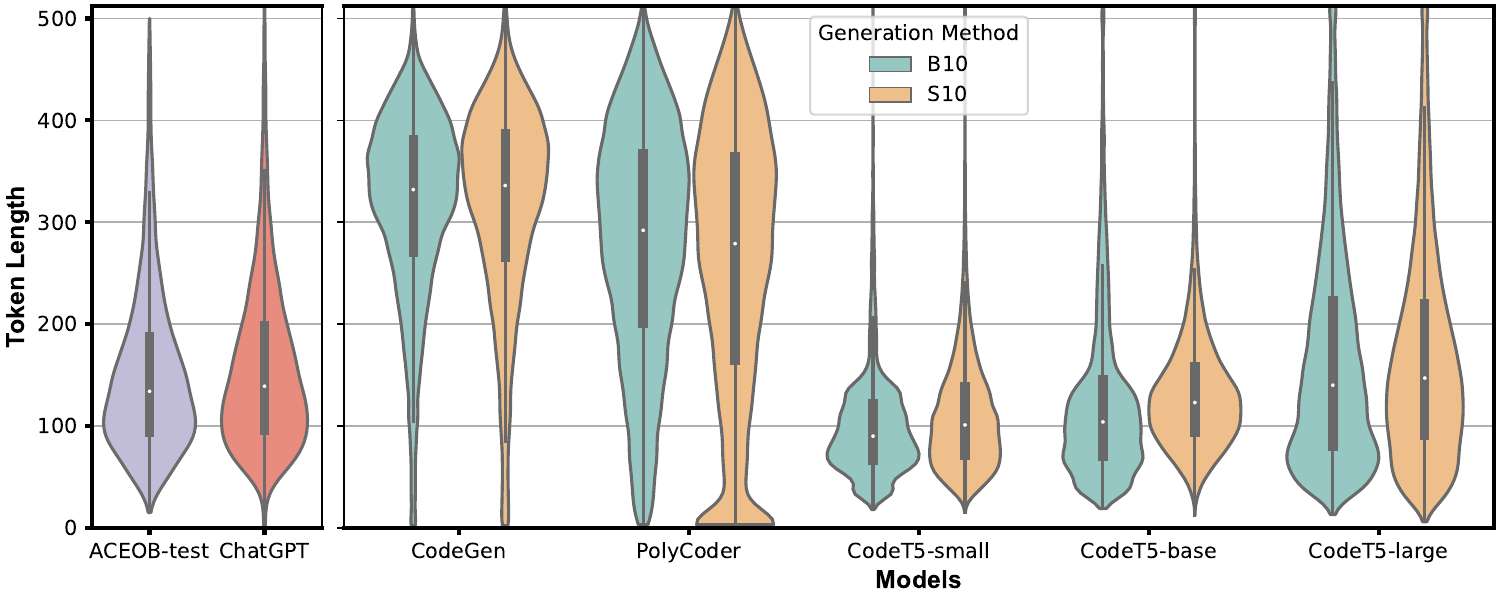}
\end{center}
\caption{Distribution of token length for approximately 940,000 lines of code generated by different models under various beam and sampling settings.}
\label{fig-Token-Violin}
\end{figure}

Finally, when comparing decoder models (such as CodeGen and PolyCoder) with encoder-decoder models (such as the CodeT5 series), the latter clearly performs better in IC2EC tasks. Figure \ref{fig-NPI-Violin} shows the distribution of NPI scores for codes generated by different models. Encoder-decoder models tend to generate more efficient codes, while decoder models tend to generate less efficient codes. Furthermore, Figure \ref{fig-Token-Violin} shows the distribution of code token lengths generated by different models under beam and sampling settings, where encoder-decoder models produce code token lengths closer to the real value distribution of the ACEOB test set. This suggests that encoder-decoder models may have higher performance in IC2EC tasks.

\begin{center}
\begin{tcolorbox}[colback=gray!10,
                  colframe=black,
                  width=16.2cm,
                  arc=1mm, auto outer arc, 
                  boxrule=0.5pt, %
                  toprule=1.2pt, rightrule=1.2pt, bottomrule=1.2pt,leftrule=1.2pt, 
                  ]
                  
In summary, we conclude the following:
\begin{itemize}
\item  Using the NPI filter significantly enhances model performance.
\item  Encoder-decoder models perform better in IC2EC tasks.
\item  ChatGPT excels in I/O metrics but needs improvement in learning code efficiency.
\end{itemize}

\end{tcolorbox}
\end{center}

\subsection{RQ3: Distinctive Domains of Beam Search and Sampling Expertise}\label{8.6}

From the data analysis in Tables \ref{tab-Overall-decoder} and \ref{tab-Overall-en-decoder}, we noticed several significant trends. Even without applying the NPI filter, greedy search (beam search with a width of 1) still outperformed random sampling in terms of NPI scores. However, when the NPI filter was applied, the optimization effect of the sampling method significantly exceeded that of beam search, indicating that sampling methods heavily rely on effective code filtering to rank code quality.

Further analyzing the IOCCB metric, we found that sampling methods (such as S5 and S10) slightly outperformed beam search (like B5 and B10) in most cases. This trend persisted even after optimizing with the NPI filter, indicating that although sampling increased code diversity, it did not significantly affect the token similarity of the code to the desired output. Figure \ref{fig_B10_S10_NPI} more intuitively demonstrated that in S10 configurations across various models, sampling methods usually outperformed B10 configurations.

We also explored why the NPI filter optimization effect exceeded that of beam search. Table \ref{tab-Similarity-CodeBLEU} compared the average CodeBLEU scores between codes generated under beam (B5 and B10) and sampling (S5 and S10) searches. Specifically, CodeGen's B5 represents the average CodeBLEU score among codes generated at a beam width of 5, averaged over all test set data. The CodeBLEU scores for beam search were significantly higher than for sampling, implying that they lacked diversity. Although increasing beam width could increase diversity, it also added to computational complexity. Furthermore, beam search might exclude potential optimal solution candidates in its early stages, thus lowering the possibility of finding the global optimum—a characteristic known as the “greedy nature of beam search”.

\begin{table}[htbp]
\caption{Similarity (CodeBLEU scores) between generated codes by five models using beam search with beam sizes of 5 and 10, and by sampling. A \textbf{lower} score indicates a \textbf{higher} degree of diversity.}
\begin{center}
\begin{tabularx}{0.9\textwidth}{X X X X r}
\hline
\textbf{Models}     &     \textbf{\textit{B5}}   &    \textbf{\textit{B10}}   &   \textbf{\textit{S}}   &   \textbf{\textit{Average}}       \\
\hline 
CodeGen           &     97.53               &   95.81                &    \textbf{3.42}    &  65.59     \\
PolyCoder         &     95.83               &   93.49                &   9.9                 &  66.41  \\
CodeT5-small      &      \textbf{84.55}     &    \textbf{82.49}      &   18.67              &  61.9  \\
CodeT5-base       &     87.3                &   83.13                &   13.61           &   \textbf{61.35}  \\
CodeT5-large      &     86.75               &   83.49                &   15.65              &  61.96 \\
\hline
\rowcolor{gray!15}  \textbf{Average}           &    \textbf{90.39}    &   \textbf{87.68}     &   \textbf{12.25}    &    \textbf{63.44}    \\
\hline
\end{tabularx}
\label{tab-Similarity-CodeBLEU}
\end{center}
\end{table}

\begin{center}
\begin{tcolorbox}[colback=gray!10,
                  colframe=black,
                  width=16.2cm,
                  arc=1mm, auto outer arc, 
                  boxrule=0.5pt, %
                  toprule=1.2pt, rightrule=1.2pt, bottomrule=1.2pt,leftrule=1.2pt, 
                  ]
                  
Overall, with the support of the NPI filter, sampling generally outperformed beam search. However, without applying the NPI filter, the situation was reversed. This suggests that due to its high diversity, sampling can achieve better performance with effective filter support, while beam search remains more stable without additional optimization measures.

\end{tcolorbox}
\end{center}

\subsection{RQ4: Impact of Difficulty on Performance}\label{8.7}

We conducted a detailed analysis of the difficulty levels in the ACEOB dataset, categorizing code complexity based on problem difficulty from the Codeforces website. In the ACEOB test set, 19 specific difficulty categories were grouped into three broader categories: Introductory (0), Interview (1-3), and Competition (4-18). The number of samples included in each difficulty level were 5372 (Introductory), 2178 (Interview), and 1755 (Competition), respectively.

Analyzing the data from Tables \ref{tab-decode-difficulty} and \ref{tab-en-decode-difficulty}, we found that as difficulty increased, the I/O metric generally decreased, indicating that more complex problems negatively affected the syntactic correctness of code. Notably, this trend was more pronounced in the ChatGPT model, suggesting that as model performance increased, the impact of complexity on code correctness became more evident.

\begin{table}[thbp]
\caption{Performance of the decoder models CodeGen and PolyCoder at different difficulty levels on the ACEOB test set.}
\begin{center}
\resizebox{0.9\textwidth}{!}{%
\begin{tabular}{c c c c c c c c c c} 
\hline
                    &   \multicolumn{3}{c}{\textbf{I/O} (\%)}
                    &   \multicolumn{3}{c}{\textbf{NPI}}  
                    &   \multicolumn{3}{c}{\textbf{IOCCB}}     \\
\cline{2-10} 

\textbf{Models}    
&    \rotatebox{80}{\textbf{\textit{Introductory}}}    &    \rotatebox{80}{\textbf{\textit{Interview}}}   &    \rotatebox{80}{\textbf{\textit{Competition}}} 
&    \rotatebox{80}{\textbf{\textit{Introductory}}}    &    \rotatebox{80}{\textbf{\textit{Interview}}}   &    \rotatebox{80}{\textbf{\textit{Competition}}} 
&    \rotatebox{80}{\textbf{\textit{Introductory}}}    &    \rotatebox{80}{\textbf{\textit{Interview}}}   &    \rotatebox{80}{\textbf{\textit{Competition}}} 
\\
\hline
CodeGen(G)  &  0  &  0  &  0  &  30.68  &  49.72  &  53.78  &  8.89  &  5.53  &  5.08   \\
\hline
B5  &  0  &  0  &  0  &  29.83  &  48.83  &  53.4  &  8.08  &  5.08  &  4.45   \\
B5+NPI  &  0  &  0  &  0  &  31.61  &  52.51  &  55.82  &  8.09  &  5.05  &  4.45   \\
S5  &  0  &  0  &  0  &  30.61  &  48.28  &  54.07  &  8.95  &  5.67  &  4.98   \\
S5+NPI  &  0  &  0  &  0  &  47.82  &  74.26  &  75.05  &  9.38  &  6.12  &  5.5   \\
\hline
B10  &  0  &  0  &  0  &  31.55  &  52.04  &  56.63  &  7.83  &  5.03  &  4.68   \\
B10+NPI  &  0  &  0  &  0  &  34.49  &  56.17  &  60.93  &  7.84  &  5.04  &  4.65   \\
S10  &  0  &  0  &  0  &  30.23  &  48.95  &  53.17  &  8.88  &  5.68  &  5.08   \\
S10+NPI  &  0  &  0  &  0  &  54.62  &  81.09  &  80.44  &  9.73  &  6.39  &  5.73   \\
\hline
\rowcolor{gray!15}    PolyCoder(G)  &  0  &  0  &  0  &  36.87  &  59.83  &  66.92  &  \textbf{10.57}  &  \textbf{8.09}  &  \textbf{9.89}   \\
\hline
\rowcolor{gray!15}   B5  &  0  &  0  &  0  &  35.81  &  57.59  &  61.58  &  8.28  &  5.94  &  5.97   \\
\rowcolor{gray!15}   B5+NPI  &  0  &  0  &  0  &  38.32  &  61.37  &  65.08  &  8.27  &  5.9  &  5.9   \\
\rowcolor{gray!15}   S5  &  0  &  0  &  0  &  36.58  &  60.42  &  66.13  &  10.45  &  8.02  &  9.83   \\
\rowcolor{gray!15}   S5+NPI  &  0  &  0  &  0  &  54.46  &  81.8  &  80.59  &  10.53  &  8.06  &  9.86   \\
\hline
\rowcolor{gray!15}   B10  &  0  &  0  &  0  &  37.21  &  59.34  &  61.91  &  8.31  &  5.98  &  6.09   \\
\rowcolor{gray!15}   B10+NPI  &  0  &  0  &  0  &  41.11  &  65.33  &  67.38  &  8.29  &  5.9  &  6.05   \\
\rowcolor{gray!15}   S10  &  0  &  0  &  0  &  36.7  &  60.17  &  66.17  &  10.49  &  8.08  &  9.80   \\
\rowcolor{gray!15}   S10+NPI  &  0  &  0  &  0  &  \textbf{61.65}  &  \textbf{87.48}  &  \textbf{84.25}  &  10.55  &  8.05  &  9.81   \\
\hline
\textbf{Average}     &  0   &  0   &  0   &  \textbf{38.9}   &  \textbf{61.4}   &  \textbf{64.63}   &  \textbf{9.08}   &  \textbf{6.31}   &  \textbf{6.54}   \\
\hline
\end{tabular}
}
\label{tab-decode-difficulty}
\end{center}
\end{table}


\begin{table}[thbp]
\caption{Performance of the encoder-decoder models CodeT5 and ChatGPT at different difficulty levels on the ACEOB test set.}
\begin{center}
\resizebox{0.9\textwidth}{!}{%
\begin{tabular}{c c c c c c c c c c} 
\hline
                    &   \multicolumn{3}{c}{\textbf{I/O} (\%)}
                    &   \multicolumn{3}{c}{\textbf{NPI}}  
                    &   \multicolumn{3}{c}{\textbf{IOCCB}}     \\
\cline{2-10} 

\textbf{Models}    
&    \rotatebox{80}{\textbf{\textit{Introductory}}}    &    \rotatebox{80}{\textbf{\textit{Interview}}}   &    \rotatebox{80}{\textbf{\textit{Competition}}} 
&    \rotatebox{80}{\textbf{\textit{Introductory}}}    &    \rotatebox{80}{\textbf{\textit{Interview}}}   &    \rotatebox{80}{\textbf{\textit{Competition}}} 
&    \rotatebox{80}{\textbf{\textit{Introductory}}}    &    \rotatebox{80}{\textbf{\textit{Interview}}}   &    \rotatebox{80}{\textbf{\textit{Competition}}} 
\\
\hline
CodeT5-small(G)  &  0.06  &  0  &  0  &  53.33  &  70.86  &  67.29  &  15.49  &  10.45  &  8.27   \\
\hline
B5  &  0.26  &  0  &  0  &  51.77  &  70.55  &  65.83  &  17.09  &  11.82  &  9.54   \\
B5+NPI  &  0.2  &  0  &  0.06  &  62.32  &  80.38  &  74.35  &  16.96  &  11.72  &  9.29   \\
S5  &  0.04  &  0  &  0  &  50.74  &  67.68  &  64.44  &  15.08  &  10.18  &  7.66   \\
S5+NPI  &  0.09  &  0  &  0  &  72.97  &  91.64  &  84.84  &  16.06  &  11.04  &  8.78   \\
\hline
B10  &  0.69  &  0  &  0.34  &  51.54  &  68.42  &  67.81  &  19.91  &  13.69  &  10.73   \\
B10+NPI  &  0.26  &  0.05  &  0.17  &  65.04  &  82.96  &  78.0  &  19.0  &  13.41  &  10.24   \\
S10  &  0.02  &  0  &  0  &  50.45  &  67.94  &  64.84  &  15.02  &  9.98  &  7.58   \\
S10+NPI  &  0.09  &  0  &  0  &  \textbf{79.61}  &  \textbf{94.43}  &  \textbf{88.39}  &  16.27  &  11.2  &  9.24   \\
\hline
\rowcolor{gray!15}    CodeT5-base(G)  &  0.22  &  0  &  0  &  48.41  &  63.49  &  62.2  &  13.29  &  8.44  &  6.74   \\
\hline
\rowcolor{gray!15}    B5  &  0.73  &  0  &  0  &  48.74  &  66.19  &  62.88  &  15.89  &  10.0  &  7.25   \\
\rowcolor{gray!15}    B5+NPI  &  0.56  &  0  &  0  &  57.53  &  75.6  &  70.26  &  15.72  &  9.89  &  7.12   \\
\rowcolor{gray!15}    S5  &  0.3  &  0  &  0  &  46.63  &  61.55  &  61.56  &  13.29  &  8.18  &  6.44   \\
\rowcolor{gray!15}    S5+NPI  &  0.3  &  0  &  0  &  69.32  &  88.57  &  82.61  &  13.57  &  8.55  &  7.03   \\
\hline
\rowcolor{gray!15}    B10  &  1.34  &  0.18  &  0.17  &  49.52  &  66.93  &  64.89  &  19.27  &  13.5  &  9.21   \\
\rowcolor{gray!15}    B10+NPI  &  0.86  &  0.14  &  0.11  &  62.23  &  79.72  &  75.12  &  18.41  &  13.11  &  8.9   \\
\rowcolor{gray!15}    S10  &  0.24  &  0  &  0  &  46.71  &  62.35  &  61.71  &  13.21  &  8.2  &  6.31   \\
\rowcolor{gray!15}    S10+NPI  &  0.2  &  0  &  0  &  76.26  &  93.12  &  86.97  &  13.68  &  8.77  &  7.05   \\
\hline
CodeT5-large(G)  &  0.52  &  0  &  0  &  47.44  &  60.82  &  61.4  &  15.07  &  8.56  &  6.61   \\
\hline
B5  &  1.34  &  0.05  &  0  &  47.52  &  60.78  &  58.95  &  20.04  &  11.85  &  8.02   \\
B5+NPI  &  1.14  &  0.05  &  0  &  55.84  &  69.24  &  65.99  &  19.43  &  11.62  &  7.66   \\
S5  &  0.43  &  0  &  0  &  47.58  &  60.75  &  60.61  &  16.24  &  9.56  &  7.4   \\
S5+NPI  &  0.41  &  0  &  0  &  71.04  &  89.13  &  84.2  &  19.25  &  12.27  &  9.8   \\
\hline
B10  &  1.62  &  0.05  &  0  &  48.77  &  64.07  &  62.83  & 21.81  &  14.39  &  9.0   \\
B10+NPI  &  1.14  &  0.14  &  0.06  &  60.11  &  75.76  &  72.41  &  21.31  &  13.84  &  8.66   \\
S10  &  0.5  &  0  &  0  &  47.54  &  61.82  &  61.83  &  16.5  &  9.62  &  7.66   \\
S10+NPI  &  0.35  &  0  &  0  &  79.16  &  93.86  &  88.28  &  21.38  &  13.96  &  11.29   \\
\hline
\hline
\rowcolor{gray!15}    \textbf{Average}     &  \textbf{0.52}  &  \textbf{0.02}  &  \textbf{0.03}  &  \textbf{57.34}  &  \textbf{73.65}  &  \textbf{70.39}  &  \textbf{16.97}  &  \textbf{11.03}  &  \textbf{8.28}  \\
\hline
\textbf{ChatGPT(G)}   &  \textbf{63.98}  &  \textbf{54.45}  &  \textbf{52.76}  &  54.14  &  58.8  &  52.15  &   \textbf{37.11}  &  \textbf{26.28}  &  \textbf{17.42}  \\
\hline

\end{tabular}
}
\label{tab-en-decode-difficulty}
\end{center}
\end{table}


For the NPI metric, we unexpectedly found that NPI scores tended to increase with the problem difficulty. We speculate this might be due to more complex problems requiring more complex code implementations, thereby providing more opportunities for simple optimizations.

Regarding the IOCCB metric, we observed a decline with increasing problem difficulty, suggesting that problem complexity to some extent affected token similarity in code optimization tasks.

\begin{center}
\begin{tcolorbox}[colback=gray!10,
                  colframe=black,
                  width=16.2cm,
                  arc=1mm, auto outer arc, 
                  boxrule=0.5pt, %
                  toprule=1.2pt, rightrule=1.2pt, bottomrule=1.2pt,leftrule=1.2pt, 
                  ]
                  
In summary, as problem difficulty increased, I/O and IOCCB metrics showed a downward trend, while NPI metrics exhibited an opposite upward trend. This phenomenon indicates that problem difficulty significantly affects the functionality and token similarity of code, and more complex problems may entail greater optimization potential.

\end{tcolorbox}
\end{center}

\subsection{RQ5: Impact of Algorithmic Tags on Performance}\label{8.8}

Upon detailed statistics of algorithmic tags in the ACEOB dataset, we found that the dataset averaged 2.25 algorithmic tags per entry. Among 9305 entries, “Greedy” (56\%), “Implement” (41\%), and “Math” (32\%) were the most common algorithmic tags, accounting for 91\%. We analyzed the performance of various models under different algorithmic labels through Tables \ref{tab-decoder-tag} and \ref{tab-en-decoder-tag}.

\begin{table}[thbp]
\caption{Performance of the decoder models CodeGen and PolyCoder under the “Greedy”, “Implement”, and “Math” algorithm tags in the ACEOB test set.}
\begin{center}
\resizebox{0.9\textwidth}{!}{%
\begin{tabular}{c c c c c c c c c c} 
\hline
                    &   \multicolumn{3}{c}{\textbf{I/O} (\%)} 
                    &   \multicolumn{3}{c}{\textbf{NPI}}  
                    &   \multicolumn{3}{c}{\textbf{IOCCB}}     \\
\cline{2-10} 

\textbf{Models}    
&    \rotatebox{80}{\textbf{\textit{Greedy}}}    &    \rotatebox{80}{\textbf{\textit{Implement}}}   &    \rotatebox{80}{\textbf{\textit{Math}}} 
&    \rotatebox{80}{\textbf{\textit{Greedy}}}    &    \rotatebox{80}{\textbf{\textit{Implement}}}   &    \rotatebox{80}{\textbf{\textit{Math}}} 
&    \rotatebox{80}{\textbf{\textit{Greedy}}}    &    \rotatebox{80}{\textbf{\textit{Implement}}}   &    \rotatebox{80}{\textbf{\textit{Math}}} 
\\
\hline
CodeGen(G)  &  0  &  0  &  0  &  39.39  &  36.82  &  39.65  &  6.87  &  7.94  &  6.46   \\
\hline
B5  &  0  &  0  &  0  &  38.23  &  36.06  &  38.59  &  6.25  &  7.16  &  5.88   \\
B5+NPI  &  0  &  0  &  0  &  40.66  &  38.44  &  41.07  &  6.24  &  7.16  &  5.88   \\
S5  &  0  &  0  &  0  &  38.74  &  36.66  &  39.79  &  6.96  &  8.0  &  6.57   \\
S5+NPI  &  0  &  0  &  0  &  59.32  &  57.04  &  59.18  &  7.35  &  8.61  &  6.81   \\
\hline
B10  &  0  &  0  &  0  &  40.64  &  38.71  &  41.24  &  6.07  &  7.14  &  5.82   \\
B10+NPI  &  0  &  0  &  0  &  44.32  &  42.31  &  44.75  &  6.07  &  7.17  &  5.83   \\
S10  &  0  &  0  &  0  &  39.2  &  37.11  &  39.82  &  6.96  &  8.04  &  6.48   \\
S10+NPI  &  0  &  0  &  0  &  65.71  &  64.12  &  65.58  &  7.69  &  8.89  &  7.25   \\
\hline
\rowcolor{gray!15}    PolyCoder(G)  &  0  &  0  &  0  &  47.43  &  46.23  &  47.55  &  \textbf{9.45}  &  \textbf{10.75}  &  8.6   \\
\hline
\rowcolor{gray!15}    B5  &  0  &  0  &  0  &  45.08  &  42.47  &  46.11  &  6.9  &  8.05  &  6.4   \\
\rowcolor{gray!15}    B5+NPI  &  0  &  0  &  0  &  48.19  &  45.61  &  49.16  &  6.87  &  8.0  &  6.37   \\
\rowcolor{gray!15}    S5  &  0  &  0  &  0  &  47.54  &  46.23  &  47.7  &  9.44  &  10.56  &  8.46   \\
\rowcolor{gray!15}    S5+NPI  &  0  &  0  &  0  &  65.9  &  64.5  &  65.8  &  9.43  &  10.72  &  \textbf{8.62}   \\
\hline
\rowcolor{gray!15}    B10  &  0  &  0  &  0  &  46.35  &  43.22  &  47.13  &  6.93  &  8.18  &  6.34   \\
\rowcolor{gray!15}    B10+NPI  &  0  &  0  &  0  &  51.15  &  48.26  &  51.52  &  6.91  &  8.11  &  6.31   \\
\rowcolor{gray!15}    S10  &  0  &  0  &  0  &  47.48  &  46.22  &  47.86  &  9.41  &  10.62  &  8.49   \\
\rowcolor{gray!15}    S10+NPI  &  0  &  0  &  0  &  \textbf{72.05}  &  \textbf{71.22}  &  \textbf{71.87}  &  9.4  &  10.71  &  8.57   \\
\hline
\textbf{Average}    &  0  &  0  &  0  &  \textbf{48.74}  &  \textbf{46.74}  &  \textbf{49.13}  &  \textbf{7.51}  &  \textbf{8.66}  &  \textbf{6.95}   \\
\hline
\end{tabular}
}
\label{tab-decoder-tag}
\end{center}
\end{table}


\begin{table}[thbp]
\caption{Performance of the encoder-decoder models CodeT5 and ChatGPT under the “Greedy”, “Implement”, and “Math” algorithm tags in the ACEOB test set.}
\begin{center}
\resizebox{0.9\textwidth}{!}{%
\begin{tabular}{c c c c c c c c c c} 
\hline
                    &   \multicolumn{3}{c}{\textbf{I/O} (\%)}
                    &   \multicolumn{3}{c}{\textbf{NPI}}  
                    &   \multicolumn{3}{c}{\textbf{IOCCB}}     \\
\cline{2-10} 

\textbf{Models}    
&    \rotatebox{80}{\textbf{\textit{Greedy}}}    &    \rotatebox{80}{\textbf{\textit{Implement}}}   &    \rotatebox{80}{\textbf{\textit{Math}}} 
&    \rotatebox{80}{\textbf{\textit{Greedy}}}    &    \rotatebox{80}{\textbf{\textit{Implement}}}   &    \rotatebox{80}{\textbf{\textit{Math}}} 
&    \rotatebox{80}{\textbf{\textit{Greedy}}}    &    \rotatebox{80}{\textbf{\textit{Implement}}}   &    \rotatebox{80}{\textbf{\textit{Math}}} 
\\
\hline
CodeT5-small(G)  &  0.02  &  0.06  &  0.07  &  59.54  &  59.04  &  61.31  &  12.3  &  12.99  &  12.89   \\
\hline
B5  &  0.1  &  0.14  &  0.1  &  57.95  &  57.3  &  59.35  &  13.17  &  14.39  &  13.99   \\
B5+NPI  &  0.06  &  0.17  &  0.07  &  68.49  &  67.58  &  69.47  &  13.18  &  14.25  &  13.85   \\
S5  &  0.04  &  0.06  &  0.03  &  57.4  &  56.26  &  58.49  &  11.98  &  12.87  &  12.32   \\
S5+NPI  &  0.02  &  0.06  &  0.07  &  80.26  &  79.41  &  80.33  &  12.79  &  13.52  &  13.4   \\
\hline
B10  &  0.4  &  0.55  &  0.46  &  57.6  &  57.79  &  59.01  &  15.44  &  16.18  &  16.4   \\
B10+NPI  &  0.1  &  0.33  &  0.03  &  71.58  &  71.41  &  72.32  &  14.74  &  15.69  &  15.54   \\
S10  &  0.02  &  0.14  &  0.07  &  56.61  &  56.56  &  59.07  &  11.88  &  12.73  &  12.17   \\
S10+NPI  &  0  &  0.03  &  0  &  \textbf{85.78}  &  85.4  &  \textbf{84.89}  &  13.07  &  13.81  &  13.71   \\
\hline
\rowcolor{gray!15}    CodeT5-base(G)  &  0.04  &  0.22  &  0.07  &  53.09  &  53.88  &  55.41  &  10.33  &  11.3  &  10.33   \\
\hline
\rowcolor{gray!15}    B5  &  0.25  &  0.61  &  0.4  &  54.47  &  53.57  &  56.61  &  11.97  &  12.8  &  12.85   \\
\rowcolor{gray!15}    B5+NPI  &  0.19  &  0.44  &  0.33  &  63.58  &  62.88  &  65.15  &  11.8  &  12.65  &  12.67   \\
\rowcolor{gray!15}    S5  &  0.04  &  0.33  &  0.07  &  52.64  &  53.18  &  54.51  &  10.17  &  11.37  &  10.11   \\
\rowcolor{gray!15}    S5+NPI  &  0.06  &  0.19  &  0.1  &  76.53  &  76.76  &  76.98  &  10.57  &  11.58  &  10.57   \\
\hline
\rowcolor{gray!15}    B10  &  0.59  &  1.18  &  0.76  &  55.51  &  54.79  &  57.3  &  15.33  &  15.68  &  15.67   \\
\rowcolor{gray!15}    B10+NPI  &  0.34  &  0.74  &  0.4  &  68.54  &  68.07  &  69.81  &  14.71  &  15.06  &  14.92   \\
\rowcolor{gray!15}    S10  &  0.06  &  0.19  &  0.1  &  52.14  &  51.94  &  54.44  &  10.24  &  11.3  &  10.24   \\
\rowcolor{gray!15}    S10+NPI  &  0.04  &  0.08  &  0.07  &  82.88  &  83.24  &  82.49  &  10.67  &  11.7  &  10.67   \\
\hline
CodeT5-large(G)  &  0.19  &  0.39  &  0.23  &  52.27  &  52.36  &  54.1  &  11.24  &  12.45  &  11.56   \\
\hline
B5  &  0.63  &  0.72  &  1.0  &  51.42  &  51.43  &  53.13  &  15.24  &  16.6  &  15.15   \\
B5+NPI  &  0.61  &  0.5  &  0.86  &  59.93  &  60.25  &  61.36  &  14.95  &  15.86  &  14.75   \\
S5  &  0.17  &  0.28  &  0.17  &  52.41  &  52.92  &  54.03  &  12.33  &  13.89  &  12.34   \\
S5+NPI  &  0.21  &  0.22  &  0.33  &  78.09  &  78.17  &  78.19  &  15.45  &  16.71  &  14.87   \\
\hline
B10  &  0.78  &  0.96  &  1.23  &  53.6  &  54.19  &  55.54  &  16.97  &  18.69  &  16.79   \\
B10+NPI  &  0.57  &  0.66  &  0.83  &  65.38  &  65.51  &  67.24  &  16.5  &  18.26  &  16.22   \\
S10  &  0.19  &  0.3  &  0.23  &  52.76  &  52.56  &  53.62  &  12.64  &  13.75  &  12.65   \\
S10+NPI  &  0.17  &  0.17  &  0.3  &  85.09  &  \textbf{85.44}  &  84.33  &  17.39  &  18.68  &  16.68   \\
\hline
\hline
\rowcolor{gray!15}    \textbf{Average}    &  0.22  &  0.36  &  0.31  &  \textbf{63.17}  &  \textbf{63.03}  &  \textbf{64.39}  &  \textbf{13.22}  &  \textbf{14.25}  &  \textbf{13.46}   \\
\hline
\textbf{ChatGPT(G)}   &  \textbf{58.81}  &  \textbf{58.16}  &  \textbf{62.0}  &  55.44  &  55.16  &  54.38  &  \textbf{30.84}  &  \textbf{31.63}  &  \textbf{29.18}   \\
\hline

\end{tabular}
}
\label{tab-en-decoder-tag}
\end{center}
\end{table}


For the I/O metric, CodeT5 performed best under the “Implement” label, while ChatGPT was the opposite. For the NPI metric, all models generally performed poorly under the “Implement” label, while for the IOCCB metric, all models showed better results.

\begin{center}
\begin{tcolorbox}[colback=gray!10,
                  colframe=black,
                  width=16.2cm,
                  arc=1mm, auto outer arc, 
                  boxrule=0.5pt, %
                  toprule=1.2pt, rightrule=1.2pt, bottomrule=1.2pt,leftrule=1.2pt, 
                  ]

Overall, while the impact of algorithmic labels on performance was relatively subtle, data under the “Implement” label exhibited clear specificity. This specificity was quite similar to the performance under high problem difficulty, possibly because these problems inherently have a higher level of difficulty. This indicates that although labels point to specific knowledge domains, LLMs may exhibit special or extreme performance in specific categories, like “Implement”. These findings might point to a potential correlation between algorithmic labels and problem difficulty, warranting further exploration.

\end{tcolorbox}
\end{center}

\subsection{RQ6: Performance of ChatGPT}\label{8.9}

\begin{figure}[thbp]
\begin{center}
\includegraphics[width=0.7\textwidth]{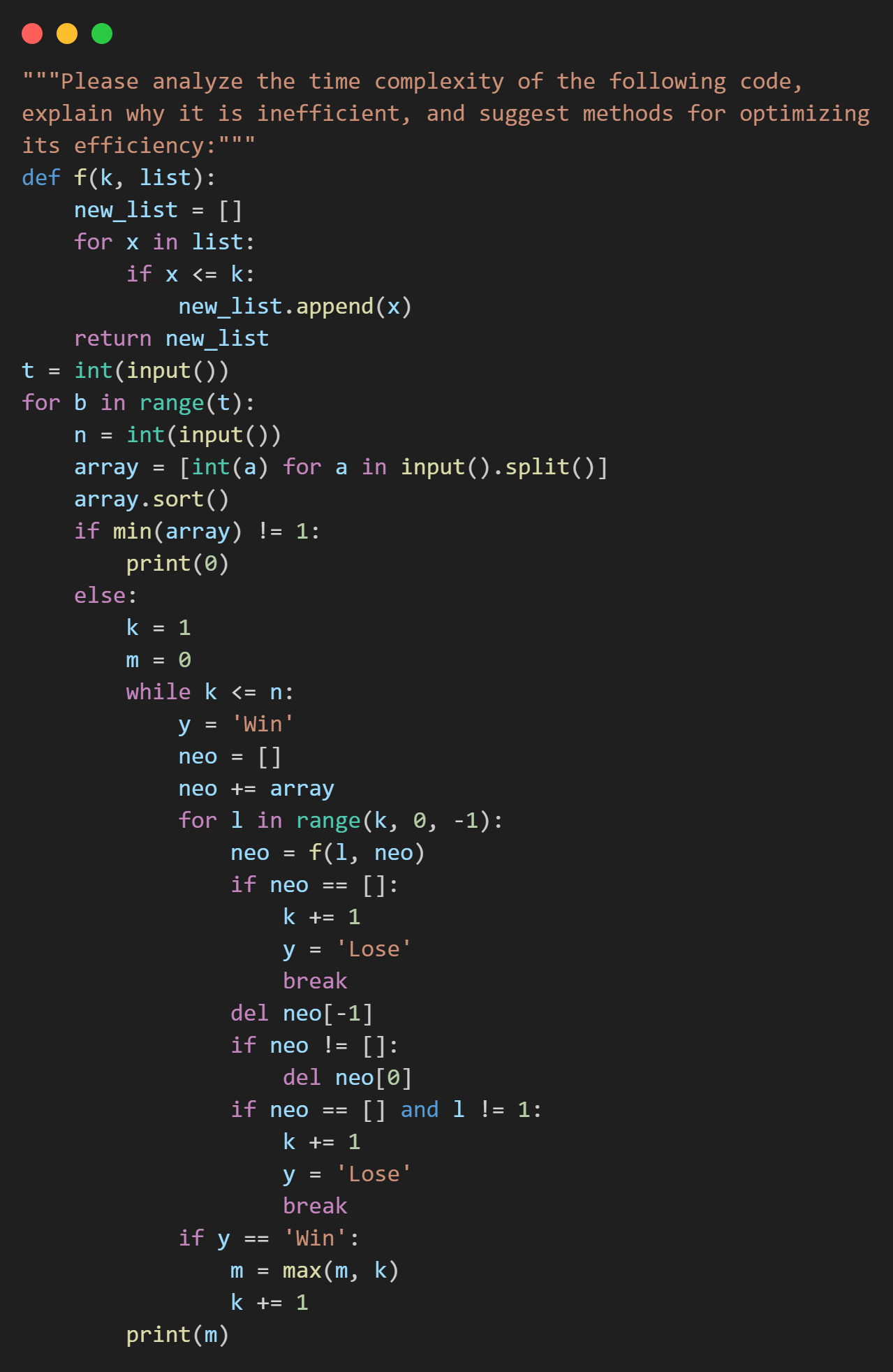}
\end{center}
\caption{ChatGPT's input examples for analyzing the time complexity of inefficient code, reasons for inefficiency, and efficiency optimization strategies.}
\label{fig-ChatGPT-input}
\end{figure}

\begin{figure}[thbp]
\begin{center}
\includegraphics[width=0.8\textwidth]{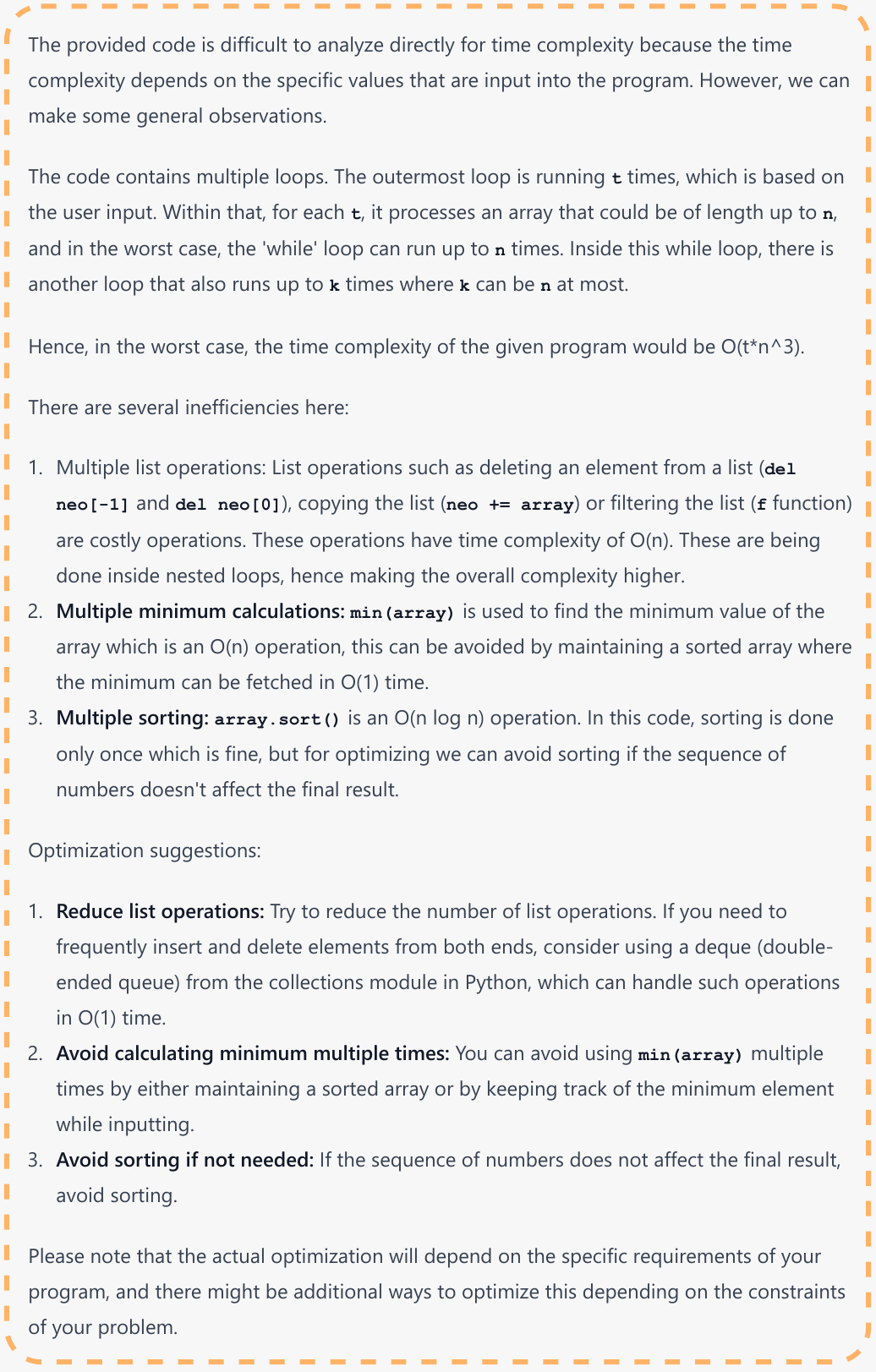}
\end{center}
\caption{The response results reveal how ChatGPT analyzes the time complexity of the code, identifies the causes of code inefficiency, and proposes effective efficiency optimization strategies.}
\label{fig-ChatGPT-response}
\end{figure}

We introduced ChatGPT into the IC2EC task to assess its performance in learning code efficiency. As a leading language model, ChatGPT's capabilities in analyzing time complexity of inefficient codes, investigating the causes of inefficiencies, and proposing efficiency optimization strategies are particularly crucial. As shown in Figure \ref{fig-ChatGPT-input}, we prompted ChatGPT to analyze specific cases of inefficient code and propose improvement strategies. Figure \ref{fig-ChatGPT-response} provides specific examples of ChatGPT's responses.

From ChatGPT's responses, we observed its accuracy and systematic approach in understanding the time complexity of code, diagnosing efficiency bottlenecks, and proposing optimization strategies. Moreover, according to the data in Table \ref{tab-Overall-en-decoder}, codes generated by ChatGPT successfully passed all the I/O tests at a rate of 59.69\%, indicating its proficiency in grasping code syntax rules.

However, despite ChatGPT's apparent advantages in theoretical knowledge application, including analyzing code time complexity, exploring reasons for inefficiencies, and proposing optimization strategies, its performance in generating efficient code for the IC2EC task has not fully met expectations. This discrepancy between its profound theoretical knowledge and its actual code optimization highlights the challenges in practical code optimization.

We analyze that while OpenAI's reports indicate that ChatGPT has been exposed to a vast amount of data, including code efficiency analysis, the lack of a specialized dataset for code efficiency optimization might be one of the reasons for its subpar performance in practical tasks. Currently, ChatGPT is capable of precise theoretical analysis, but its ability in practical code efficiency optimization still needs enhancement.

\begin{center}
\begin{tcolorbox}[colback=gray!10,
                  colframe=black,
                  width=16.2cm,
                  arc=1mm, auto outer arc, 
                  boxrule=0.5pt, %
                  toprule=1.2pt, rightrule=1.2pt, bottomrule=1.2pt,leftrule=1.2pt, 
                  ]
                  
Overall, ChatGPT's actual performance in the domain of code efficiency has not yet reached the level of other areas. Despite excelling in syntactic rules and theoretical knowledge, its optimization effect in IC2EC tasks remains limited, which may be attributed to the lack of specialized datasets for code efficiency optimization and practical experience.

\end{tcolorbox}
\end{center}

\section{Discussion}\label{9}

\subsection{Discovering Significance}\label{9.1}

Through experimental analysis of the IC2EC task, this study has revealed several \textbf{key findings} and \textbf{their underlying reasons.} Specifically:
\begin{itemize}
\item \textbf{Applicability of the IOCCB Metric:}  Compared to the CodeBLEU metric, the IOCCB metric is more suitable for IC2EC tasks, particularly when dealing with simple code (shorter token lengths or smaller execution times).
\item \textbf{Cost Model Input:} The performance of the CodeT5 model, which directly uses source code as input, surpasses that of models handling AST-segmented data. This phenomenon may stem from the fact that the CodeT5 model was not specifically trained with AST inputs during its pre-training phase, resulting in its less effective extraction of feature information from AST.
\item \textbf{Cost Model Optimization:}  Two cost models perform best when predicting code with 250-300 token lengths. For different execution times, the NPI prediction model is more suitable for medium execution times, while the execution time model is applicable for extreme durations.
\item \textbf{Architecture Comparison:}   Encoder-decoder architecture excels in the IC2EC task, likely due to its proficiency in handling complex sequence-to-sequence tasks.
\item \textbf{Sampling Advantage:}   Sampling methods generally outperform beam search, especially when combined with NPI filters, where diversity becomes a significant strength.
\item \textbf{Dependency on Filtering Mechanism:}   Sampling methods rely heavily on an effective code filtering mechanism to differentiate between efficiencies of generated codes.
\item \textbf{Influence of Problem Difficulty:}  The difficulty of competition problems affects not only the similarity of code tokens but also the magnitude of efficiency optimization. As difficulty increases, so does the potential for code optimization.
\item \textbf{Impact of Tags:}  The impact of problem tags is weaker compared to problem difficulty, suggesting that tags are more related to knowledge domains, while difficulty directly correlates with the complexity of solutions. Notably, problems tagged as “Implement” share characteristics with high-difficulty problems, indicating a potential correlation between algorithm tags and problem difficulty.
\item \textbf{ChatGPT Performance:}   Despite its strong theoretical knowledge, ChatGPT's performance in actual IC2EC tasks is not entirely satisfactory, hinting that the absence of a specialized code efficiency optimization dataset might affect its performance.
\end{itemize}

\subsection{Self-Reflection}\label{9.2}

\begin{table}[htbp]
\caption{Analysis of the impact of computational formula variations on the correlation (Spearman rank correlation coefficient) between the IOCCB metric and execution time, including scenarios where the unification of variable and function names is skipped in cases of code compilation failure.}
\centering
\begin{tabular*}{0.7\textwidth}{@{\extracolsep{\fill}} c | c c}

\hline
\multirow{2}{*}{\textbf{Calculation Formulas}}                &              \multicolumn{2}{c}{\textbf{ACEOB Test Set}}   \\
\cdashline{2-3}
                                           &               {\textbf{\textit{Correlation}}}              &                 {\textbf{\textit{P-value}}}     \\
\hline
Skipping Normalization Process                                                               &  -0.4490                &  $e^{-324}$          \\
\hline
$B_{max}-\sqrt{(B_{avg}-A_{avg})}$                                                          &  -0.5383                      &                $e^{-324}$          \\
\hline
Normalization Process   \textbf{\&}   $B_{max}+\sqrt{(B_{avg}-A_{avg})}$                        &  \textbf{-0.6276}             &   $e^{-324}$            \\
\hline

\end{tabular*}
\label{tab-Calculation-Formulas}
\end{table}

\textbf{IOCCB Calculation Counterintuitive.} The final step of IOCCB calculation, which involves normalizing variable and function names, seems counterintuitive. Table \ref{tab-Calculation-Formulas} shows the effects of using different formulas in the final calculation step. Experiments indicate that using the highest CodeBLEU score plus the average difference is superior to the intuitive approach of subtracting the difference. This might indicate that normalization introduces a bias opposite to the maximum value bias. Our experiments validate this hypothesis by comparing arrays with and without normalization operations, looking at their average and maximum values. Post-normalization, 84.32\% of evaluation codes showed an increase in average values, while only 34.89\% showed an increase in maximum values.

\textbf{Compilation Impact.} Whether code is compilable significantly affects the accuracy of the IOCCB metric. Uncompilable code hampers the normalization of variables and function names, thus impacting the metric's precision. Table \ref{tab-Calculation-Formulas} illustrates the accuracy impact of code compilability on IOCCB. Skipping this calculation step significantly reduces the accuracy of the IOCCB metric.

\textbf{Runtime Evaluation.} In assessing runtime, we didn't strictly adhere to the gold standards within the domain but adopted a custom approach. Subsequent adjustments led us to follow the standard of running the code 30 times and taking the median to improve the accuracy of the assessment.

These reflections deepen our understanding of the importance of choice and calculation methods' accuracy and credibility in executing IC2EC tasks.

\section{Threats to Validity}\label{10}

\textbf{External Validity.} Our research focuses on the Python programming language, which may limit the generalizability of the results. Although Python is one of the widely used programming languages, our findings and methods may need to be adjusted and validated for other programming languages. Furthermore, we have analyzed the coding efficiency of humans on the Codeforces programming website, which may differ from that of other programmers.

\textbf{Internal Validity.} In the experimental design of the IC2EC task, we carefully controlled variables to ensure the accuracy of the results. Despite this, minor setting changes or randomness may still affect the experimental outcomes. To address this threat, we ensured that only one parameter changes in each experiment, other parameters remain consistent, and conducted repeated experiments to verify the stability of the results.

\section{Limitations}\label{11}

The nature of performance optimization issues is diverse, extending far beyond algorithmic complexity. The study by Böck et al. \mbox{[\citealp{bock2023performance}]} reveals that methods using programming competition datasets, even after pre-training and fine-tuning for domain-specific code, may not effectively identify performance issues in AAA game development. This finding underscores the potential limitations our ACEOB method may face when applied to different performance problems. Although ACEOB focuses on algorithmic complexity, we must also acknowledge that many performance bottlenecks stem from the complex interactions between software and I/O resources, databases, networks, and system architectures. Moreover, building ACEOB on programming competition datasets means it primarily deals with problem types and scales that may differ from those in real, large-scale software engineering projects. Programming competition problems often focus on specific algorithmic issues and are relatively small in scale, whereas actual software projects may include a broad range of functional modules and higher complexity. Therefore, although our method in its current form may not be suitable for performance optimization in large-scale software systems, we believe that optimizing the efficiency of large codebases is not only possible but also necessary with technological advancements and future research expansion.

\section{Conclusion}\label{12}

This study introduced ACEOB, a benchmark dataset focused on evaluating the performance of LLMs on the IC2EC task. Through the analysis of 95,359 efficient-inefficient code pairs, we not only focused on the functionality of code generation but also paid more attention to optimizing code efficiency. By introducing IOCCB and NPI evaluation metrics, this study provides a precise method for measuring the capability of the IC2EC task. Our evaluation results reveal that there is still room for improvement in the performance of mainstream generative models in code efficiency optimization and demonstrate the significant advantage of the NPI filter in the IC2EC task. As more emphasis is placed on the ability to optimize code efficiency, the ACEOB benchmark provides an important tool for tracking research progress in this field. In the future, with the advancement of models and technology, more research is expected to leverage this benchmark to propel the development of the efficiency optimization field.










\printcredits




\bibliographystyle{elsarticle-num}



\end{document}